\documentclass[preprint,aps,nofootinbib,11pt]{revtex4-1}
\usepackage{geometry}                % See geometry.pdf to learn the layout options. There are lots.
\usepackage{amsmath,amssymb,amsfonts,dcolumn,color,graphicx,graphics,latexsym,placeins,epsfig,tikz}
\usetikzlibrary{arrows,shapes}
\usepackage{subfigure,rotating,bm,mathrsfs}

\newcommand{\RN}[1]{
\textup{\uppercase\expandafter{\romannumeral#1}}
}
\newcommand{\be}{\begin{equation}}
\newcommand{\ee}{\end{equation}}
\newcommand{\ba}{\begin{eqnarray}}
\newcommand{\ea}{\end{eqnarray}}

\usepackage[pagebackref=false, colorlinks=true]{hyperref}
\definecolor{redish}{rgb}{0.7,0.2,0.0}  % color defined in (r=red,g=green,b=blue) model
\definecolor{bluish}{rgb}{0.2,0.5,0.8}
\hypersetup{linkcolor=redish,          % color of internal links
                  citecolor=blue,        % color of links to bibliography
                  filecolor=magenta,      % color of file links
                  urlcolor=blue}          % color of the url

\begin{document}
\title{\Large Collapse in $f(R)$ gravity and the method of $R$ matching}
\author{Sandip Chowdhury}
\email{sandipc@iitk.ac.in}
\author{Kunal Pal}
\email{kunalpal@iitk.ac.in}
\author{Kuntal Pal}
\email{kuntal@iitk.ac.in}
\author{Tapobrata Sarkar}
\email{tapo@iitk.ac.in}
\affiliation{{\it Department of Physics, Indian Institute of Technology, Kanpur 208016, India}}
%%%%%%%%%%%%%%%%%%%%%%%%%%%%%%%%%%%%%%%%%%%%%%%%
\begin{abstract}
Collapsing solutions in $f(R)$ gravity are restricted due to junction conditions that demand continuity of 
the Ricci scalar and its normal derivative across the time-like collapsing hypersurface. These are
obtained via the method of $R$-matching,
which is ubiquitous in $f(R)$ collapse scenarios. In this paper, we 
study spherically symmetric collapse with the modification term $\alpha R^2$, and use $R$-matching to exemplify a
class of new solutions. After discussing some mathematical preliminaries by which we obtain an algebraic relation between
the shear and the anisotropy in these theories, we consider two metric ansatzes. 
In the first, the collapsing metric is considered to be a separable function of the co-moving radius and time, 
and the collapse is shear-free, and in the second, a non-separable interior solution is considered, that represents 
gravitational collapse with non-zero shear viscosity. We arrive
at novel solutions that indicate the formation of black holes or locally naked singularities, while obeying all 
the necessary energy conditions. The separable case allows for a simple analytic expression of the energy-momentum tensor, 
that indicates the positivity of the pressures throughout collapse, and is further
used to study the heat flux evolution of the collapsing matter, whose analytic solutions are presented under 
certain approximations. These clearly highlight the role of modified gravity in the examples that we consider. 
\end{abstract}
\maketitle
\section{Introduction}

Einstein's general relativity (GR) is the most successful theory of gravity till date, although modifications to GR
continue to attract much attention. One of the primary reasons for attempting such modifications has to do with
explaining the late time acceleration of the universe. It is known that this phenomenon is
compatible with GR in the presence of a dark energy component in the stress tensor. However, much work has been
done over the last two decades in trying to explain cosmic acceleration of the universe 
without invoking dark forms of matter and energy. One such 
candidate theory is $f(R)$ gravity\footnote{We will always deal with metric $f(R)$ gravity in this paper and perform calculations
in the Jordan frame.}
(for a sampling of the literature, see the excellent reviews \cite{FRRev1}, 
\cite{FRRev2}, \cite{FRRev3}) obtained by modifying the Einstein-Hilbert action to one which includes
a regular function $f(R)$ of the Ricci scalar $R$, i.e one in which the Lagrangian density is $R+f(R)$, apart from
the matter part. In this paper,
we will deal with the specific model $f(R) = \alpha R^2$, with $\alpha$ being a positive constant, a model 
famously proposed in \cite{Starobinsky}. 

While phenomenological studies of $f(R)$ gravity abound in the literature, there has been relatively lesser
focus on collapse scenarios, where matter collapses under its own gravitational 
force, with the underlying theory being $f(R)$ gravity. We briefly mention a few relevant papers to highlight
the progress made thus far. 
In \cite{Bamba}, the collapse process of a star was considered in modified gravity, and it was shown that 
a class of $f(R)$ theories can result in the prevention of a central singularity in such a process. 
A generic study of collapse processes of self gravitating dust in $f(R)$ gravity was initiated in \cite{Cem}. 
In \cite{GM}, this process was studied for the case of null dust. In the context of cosmology, collapse
in modified gravity was studied in \cite{Koppetal}, while an extensive numerical analysis for black hole formation
in these theories was carried out in \cite{Guo}. A more recent analysis on collapsing stars in modified gravity 
was done in \cite{Goswami} (with a generalisation to conformally flat stars appearing in \cite{GB})
while results on the collapse of a perfect fluid in $f(R)$ gravity was reported in \cite{Chakrabarti}. 

As is well known by now, collapse situations in
$f(R)$ gravity are greatly restricted compared to their GR counterparts, due to stringent boundary conditions. In GR, 
such boundary conditions, known as the Darmois-Israel conditions \cite{Darmois},\cite{Israel} 
require the first and second fundamental forms to match on the collapsing hypersurface, which is 
a time-like junction between an internal and an external region of space-time. This guarantees smooth matching of the two
regions of space-time, i.e without a stress tensor at the junction. 
In $f(R)$ gravity, on the other hand, additional conditions have to be imposed \cite{Deruelle},\cite{Senovilla} (see also \cite{Mars})
over and above the Darmois-Israel conditions. These often require the Ricci scalar and its (normal) derivative to vanish at the boundary, for 
smooth matching of the collapsing region with an external Schwarzschild space-time. 

This fact was exploited fairly recently in \cite{Goswami} to provide some realistic models of gravitational collapse
in $f(R)$ theories in which the coefficient of viscosity is turned off. The starting point of
the analysis is the assumption of a specific form of a time dependent spherically symmetric metric, that depends on an
arbitrary function of the radial coordinate. The modified Einstein equations in $f(R)$ gravity
are then used to  constrain these functions in such a way that the extra junction conditions are satisfied, 
and specific choices give concrete examples of collapse scenarios in $f(R)$ models. Importantly, as pointed
out in \cite{Goswami}, the additional junction conditions mentioned in \cite{Deruelle},\cite{Senovilla} render a straightforward
generalisation of collapse processes in GR, to scenarios involving modified gravity, difficult. We should emphasise here
that in addition to the junction conditions, the collapsing fluid must satisfy various energy conditions that we will
elaborate upon in sequel. In totality, all this amounts to the fact that analysing collapsing scenarios in $f(R)$ gravity
might be a substantially complicated task. 

In this paper, we present new solutions for collapse in $f(R)$ gravity, by assuming some simple ansatzes for the metric,
which is then solved by the extra junction conditions, 
namely the matching of the Ricci scalar and its derivative across a
time-like boundary. This the $R$-matching method commonly used in $f(R)$ collapse scenarios.
This is elaborated upon for two cases, first when the metric consists of separable functions of the radial and the
time coordinate, and second when it is not. Importantly, the second condition admits shear, and we study this
in the presence of a non-zero coefficient of shear viscosity. 
The $R$-matching method gives us the full solution of the modified 
Einstein equations, and we are able to provide a class of realistic collapse models in $f(R)$ gravity, consistent 
with all energy conditions.
For separable solutions to the metric, we are able to provide simple analytic expressions for the components of the 
energy momentum tensor. These are then used to construct analytic solutions of the heat flux evolution equation. 

This paper is organised as follows. In the next section, after a brief review of the necessary formalism, we write down the
general evolution equation of the shear in $f(R)$ theories. 
The general equation for the evolution of shear is written down and we obtain an algebraic relation
between the shear and the anisotropy in $f(R)$ collapse models, via this formula. 
After this, the necessary energy conditions and the junction conditions
of the collapsing fluid are reviewed. With these ingredients, in section 3, we construct 
a separable solution of the metric using the $R$-matching method, and show that the end state of collapse is necessarily a black hole. 
In this case, the collapse is shear-free. Then
in section 4, we extend this to non-separable solutions and construct collapsing solutions that obey all energy conditions
with the end state being a (locally) naked singularity. The role of shear is commented upon, in this example. In section 5, we study
some physical properties of the collapsing fluid, for the separable case. 
The nature of the equation of state is commented upon, and the heat flux evolution 
equation is solved under some assumptions to clearly highlight the role of the $f(R)$ parameter. Finally, section 6 ends this
paper with a summary of the main results and some discussions. 

\section{Mathematical preliminaries and set up}

For a generic collapse scenario, in co-moving coordinates $(t,r,\theta,\phi)$ the metric inside the 
spherically symmetric collapsing cloud is written as 
\begin{equation}
ds^{2}_{-}=-e^{2\nu(r,t)}dt^{2}+e^{2\psi(r,t)}dr^{2}+Q^{2}(r,t)d\Omega^{2}~,
\label{inner}
\end{equation}
where $d\Omega^2 = d\theta^2 + \sin^2\theta d\phi^2$. 
The metric outside the collapsing matter is usually represented by the Vaidya solution in terms of the retarded time $u$ as 
\begin{equation}
ds_+^2 = -\left(1-\frac{2m(u)}{\tilde r}\right)du^2 - 2dud{\tilde r} + {\tilde r}^2d\Omega^2
\label{outer}
\end{equation}
In this paper, we will be interested in an exterior vacuum solution (i.e without any radiation) and hence with $m(u)$ being a
constant, the metric out side the collapsing matter can be taken to be the Schwarzschild metric, given by 
\begin{equation}
ds^{2}_{+}=-H(\tilde{r})d\tilde{t}^{2}+H(\tilde{r})^{-1}d\tilde{r}^{2}+\tilde{r}^{2}d\Omega^{2}~,~~
H(\tilde{r})=1-\frac{2m}{\tilde{r}} ~,
\label{outersch}
\end{equation}
where $m$ is the (constant) Schwarzschild mass, so that the heat flux obtained from eq.(\ref{inner}) is zero at the
matching hypersurface. 

The modified Einstein's equations for a Lagrangian density $R+f(R) + {\mathcal L}_{matter}$ are given by: 
\begin{equation}
G_{\mu\nu}=\frac{1}{1+F}\left(T_{\mu\nu}+D_{\mu\nu}F(R)+\frac{1}{2}g_{\mu\nu}(f-RF)\right)~,~~T_{\mu\nu} = 
-\frac{1}{\sqrt{-g}}\frac{\delta {\mathcal L}_{matter}}{\delta g^{\mu\nu}}
\label{Eeq}
\end{equation}
where
\begin{equation}
F(R)=\frac{df(R)}{dR}~,~D_{\mu\nu}=\nabla_{\mu}\nabla_{\nu}-g_{\mu\nu}\nabla_{\alpha}\nabla^{\alpha}~,~g = {\rm Det}\left[g_{\mu\nu}\right]
\end{equation}
We need to solve the modified Einstein equations with the energy momentum tensor\footnote{We work in units such that
$c$ = $8\pi G$ = $1$, with $c$ being the speed of light and $G$ is the Newton's constant.}
\begin{equation}
T_{\mu\nu}=\rho u_{\mu}u_{\nu} + P h_{\mu\nu} -\Pi_{\mu\nu} + 
2qu_{(\mu}n_{\nu)} - 2\eta\sigma_{\mu\nu}~,
\label{emt}
\end{equation}
where we define the quantities 
\begin{equation}
\sigma_{\mu\nu} = u_{(\mu;\nu)} + a_{(\mu}u_{\nu)} - \frac{1}{3}\Theta\left(g_{\mu\nu} + u_{\mu}u_{\nu}\right)~,~~
\Pi_{\mu\nu}= \Pi\left(n_{\mu}n_{\nu} - \frac{1}{3}h_{\mu\nu}\right)~,~~\Pi = p_{\theta}-p_r ~,~~
P = \frac{p_r + 2p_{\theta}}{3}~,
\label{emt1}
\end{equation}
where $(,)$ denote a symmetrization, and a semicolon denotes a covariant derivative.
Here, $\rho$ is the energy density, $p_r$ and $p_{\theta}$ are the radial and tangential pressures, respectively, 
$q^{\mu} = qn^{\mu}$ is the radial heat flow vector where $n^{\mu}$ is a unit 4-vector along the radial 
direction, and $u^{\mu}$ is the 4-velocity of the fluid. 
These satisfy $n^{\mu}n_{\mu} = 1$, $u^{\mu}u_{\mu} = -1$, $u^{\mu}q_{\mu}=0$, $u^{\mu}n_{\mu}=0$. Also, 
$\Theta = u^{\nu}_{;\nu}$ is the expansion parameter and $h_{\mu\nu} = g_{\mu\nu} + u_{\mu}u_{\nu}$ is the projection tensor. 

In this paper, we will be dealing with two situations, to be elaborated in sections 3 and 4. In the former, we will consider 
shear-free collapse, with the fluid being non-geodesic. In the latter, we will consider a geodesic fluid, but with 
non-zero shear. It will therefore be useful for us to record the relations that connect these quantities, in the $f(R)$ model
that we consider. As we will see, we are led to some useful insights here. 
 
To begin with, we record the Raychaudhuri equation, which reads (see, e.g \cite{Rippl})
\begin{equation}
\label{Ray}
u^{\alpha}\Theta_{;\alpha}+\frac{1}{3}\Theta^{2}+\frac{2}{3}\sigma^{2}-a^{\mu}_{;\mu}+\frac{1}{1+F}\bigg[-\frac{1}{2}\left(R+f(R)\right)
+\left(\frac{dF}{dR}\right)h^{\mu\nu}R_{;\mu\nu}+T_{\mu\nu}u^{\mu}u^{\nu}\bigg]=0~,
\end{equation}
where we have defined the acceleration vector $a^{\mu} = u^{\mu}_{;\nu}u^{\nu}$
and $\sigma_{\alpha\beta}\sigma^{\alpha\beta}=\frac{2}{3} \sigma^{2}$.
This equation is valid only for $f(R)$ models with $d^{2}F/dR^{2}=0$, which is the case under consideration here.\footnote{This is 
straightforwardly generalised to situations where $d^{2}F/dR^{2} \neq 0$, but the expressions are lengthy, and we will not
record them here as these will not be useful for our purpose.}
Now, we will use the identity given by \cite{Her1}
\begin{equation}
\label{H34}
u^{\beta}u_{\rho}R^{\rho}_{\alpha \beta
\mu}h^{\alpha}_{\gamma}h^{\mu}_{\nu}=h^{\alpha}_{\gamma}h^{\mu}_{\nu}\left(a_{\alpha;\mu} - u^{\beta}\sigma_{\alpha\mu;\beta}\right) +
a_{\gamma}a_{\nu}-\frac{1}{3}u^{\beta}\Theta_{;\beta}h_{\gamma\nu}-\frac{1}{9}\Theta^{2}h_{\gamma\nu}-\frac{2}{3}\Theta
\sigma_{\mu\nu}-\frac{\sigma^{2}}{3}\bigg(n_{\gamma}n_{\nu}+\frac{1}{3}h_{\gamma\nu}\bigg)~,
\end{equation}
and the well known relation between Riemann and Weyl tensors given by
\begin{equation}
R^{\mu}_{\nu\rho\sigma} = C^{\mu}_{\nu\rho\sigma}  + \frac{1}{2}\left(R^{\mu}_{\rho}g_{\nu\sigma}  - R^{\mu}_{\sigma}g_{\nu\rho}
- R_{\nu\rho}\delta^{\mu}_{\sigma} + R_{\nu\sigma}\delta^{\mu}_{\rho}\right)+\frac{R}{6}\left(\delta^{\mu}_{\sigma}g_{\nu\rho} - 
\delta^{\mu}_{\rho}g_{\nu\sigma}\right)~.
\label{RW}
\end{equation}
For $f(R)$  gravity (recall that $f(R) = \alpha R^2$ with $d^{2}F/dR^{2}=0$), eq.(\ref{RW}) can be evaluated by using 
the results derived in \cite{Rippl} and after some algebra, we obtain (with $R_{;\mu;\nu} \equiv R_{;\mu\nu}$), 
 \begin{equation}
 \label{H36}
u^{\beta}u_{\rho}R^{\rho}_{\alpha \beta \mu}h^{\alpha}_{\gamma}h^{\mu}_{\nu}=
E_{\gamma\nu}+\frac{1}{2(1+F)}\bigg[-\frac{\left(R+f(R)\right)}{3}h_{\gamma\nu}+\frac{dF}{dR}
\bigg(h_{\gamma\nu}h^{\alpha\beta}R_{;\alpha\beta}-h^{\alpha}_{\gamma}h^{\mu}_{\nu}R_{;\alpha\mu}\bigg)+
\bigg( \frac{2}{3}\rho h_{\gamma\nu}
+\Pi_{\gamma\nu}+2\eta\sigma_{\gamma\nu}\bigg)\bigg]~.
\end{equation}
This generalises a corresponding result obtained for GR in \cite{Her1}. Here, $E_{\mu\nu}$ is the electric part of the Weyl tensor,
defined as $E_{\mu\nu} = C_{\mu\nu\rho\lambda}u^{\rho}u^{\lambda}$, with the magnetic part of the Weyl tensor 
vanishing identically due to spherical symmetry. 
Then, eliminating $\rho$ from eq.(\ref{H36}) using eq.(\ref{Ray}), we obtain
 \begin{equation}
 \label{H41}
u^{\beta}u_{\rho}R^{\rho}_{\alpha \beta \mu}h^{\alpha}_{\gamma}h^{\mu}_{\nu}=
E_{\gamma\nu}+\frac{1}{2(1+F)}\bigg[\frac{dF}{dR}\bigg(\frac{1}{3}h_{\gamma\nu}h^{\alpha\beta}
R_{;\alpha\beta}-h^{\alpha}_{\gamma}h^{\mu}_{\nu}R_{;\alpha\mu}\bigg)+
{\hat P}_{\mu\nu} \bigg]
-\frac{1}{3}h_{\gamma\nu}\bigg(u^{\alpha}\Theta_{;\alpha}+\frac{1}{3}\Theta^{2}-a^{\mu}_{;\mu}+\frac{2}{3}\sigma^{2}\bigg)~,
\end{equation}
where we have defined 
\begin{equation}
E_{\mu\nu} = {\mathcal E}\left(n_{\mu}n_{\nu} - \frac{1}{3}h_{\mu\nu}\right)~,~~{\hat P}_{\mu\nu} = 
\bigg(\Pi_{\gamma\nu}+2\eta\sigma_{\gamma\nu}\bigg)
\end{equation}
Equating eq.(\ref{H41}) and eq.(\ref{H34}) we get
\begin{equation}
h^{\alpha}_{\gamma}h^{\mu}_{\nu}\left(a_{\alpha;\mu} - u^{\beta}\sigma_{\alpha\mu;\beta}\right)
+a_{\gamma}a_{\nu}-\frac{1}{3}\sigma_{\gamma\nu}\big(2\Theta+\sigma)=
E_{\gamma\nu}+\frac{1}{2(1+F)}\bigg[\frac{dF}{dR}\bigg(\frac{1}{3}h_{\gamma\nu}h^{\alpha\beta}
R_{;\alpha\beta}-h^{\alpha}_{\gamma}h^{\mu}_{\nu}R_{;\alpha\mu}\bigg)
+{\hat P}_{\mu\nu}\bigg)\bigg]
+\frac{1}{3}h_{\gamma \nu}a^{\mu}_{;\mu}.
\end{equation}
Finally, contracting with $n^{\gamma}n^{\nu}$ and denoting ${\hat P} = \left(\Pi + 2\eta\sigma\right)$,
\begin{equation}
n^{\alpha}n^{\mu}\left(a_{\alpha;\mu}-u^{\beta}\sigma_{\alpha\mu;\beta}+a_{\alpha}a_{\mu}\right)
-\frac{2}{9}\sigma(2\Theta+\sigma)=\frac{2}{3}\mathcal{E}
+\frac{1}{2(1+F)}\bigg[\frac{dF}{dR}\bigg(\frac{1}{3}h^{\alpha\mu}-n^{\alpha}n^{\mu}\bigg)R_{;\alpha\mu}+\frac{2}{3}{\hat P}\bigg]
+\frac{1}{3}a^{\mu}_{;\mu}~.
\label{finalmod}
\end{equation} 
Expanding the left hand side of eq.(\ref{finalmod}), the evolution of the shear is given by the equation
\begin{equation}
e^{-\psi}\frac{da}{dr} - \frac{2}{3}e^{-\nu}\frac{d\sigma}{dt} + a^2 -\frac{2}{9}\sigma(2\Theta+\sigma) =\frac{2}{3}\mathcal{E}
+\frac{1}{2(1+F)}\bigg[\frac{dF}{dR}\bigg(\frac{1}{3}h^{\alpha\mu}-n^{\alpha}n^{\mu}\bigg)R_{;\alpha\mu}+\frac{2}{3}{\hat P}\bigg]
+\frac{1}{3}a^{\mu}_{;\mu}~,
\label{finalmod1}
\end{equation} 
with $a=n^{\mu}a_{\mu}$. Eq.(\ref{finalmod1}) is the most general evolution equation for the shear tensor 
in $f(R)=\alpha R^2$ scenarios, with $d^2F/dR^2=0$. The GR case corresponds here to $\alpha = 0$ and has been
analysed in \cite{Her1}. We can make a few comments here. 
Now note that $\sigma$ (being computed entirely from the metric) does not depend on the 
$f(R)$ parameter $\alpha$. This means that the the term in square brackets in eq.(\ref{finalmod1}) has to be independent
of $\alpha$. For $f(R)=\alpha R^2$ theories, this can be seen to imply that
\begin{equation}
\sigma = \frac{3}{4 \eta R}\left(\frac{1}{3}h^{\alpha\mu}-n^{\alpha}n^{\mu}\right)R_{;\alpha\mu} - \frac{\Pi}{2\eta} + \frac{\left(1+
2\alpha R\right)}{4\eta R}\left(\frac{\partial \Pi}{\partial \alpha}\right)
\label{finalmod2}
\end{equation}
Eq.(\ref{finalmod2}) gives an algebraic relation between the shear and the anisotropy in the $f(R)$ theories that we 
consider.\footnote{Note that eq.(\ref{finalmod2}) holds only for non-zero $\alpha$. For $\alpha = 0$, the method of its
derivation becomes redundant.} To the best of our knowledge, eqs.(\ref{finalmod1}) and (\ref{finalmod2}) have not
appeared in the literature before, and provide useful insights into the dynamics of $f(R)$ collapse. These equations will
be identically satisfied in the explicit solutions that we will construct in sequel. 

The next ingredient in our analysis will be the relevant energy conditions of the collapsing fluid. In this context, we begin from
the energy momentum tensor of eq.(\ref{emt}), that describes the motion of a fluid with shear, with heat flow in the radial direction.
The energy conditions for such a fluid including the effects of anisotropy was obtained in \cite{Chan} (see also \cite{Chan0})
by generalising a method 
developed in \cite{Kolassis} for isotropic cases. This essentially relies on the fact that the eigenvalues of the energy momentum 
tensor should be real, and the resulting conditions on the fluid are given by 
\begin{eqnarray}
(i) &~& |\rho + p_r - 2\eta\sigma_{11}| -2|q| \geq 0~, \nonumber\\
(ii) &~& \rho - p_r + 2p_{\theta} + \Delta + 2\eta\left(\sigma_{11} - 2\sigma_{22}\right) \geq 0\nonumber\\
(iii) &~& \rho - p_r + 2p_{\theta} + \Delta + 2\eta\left(\sigma_{11} - 2\sigma_{33}\right) \geq 0
\label{con12gen}
\end{eqnarray}
where we have defined 
\begin{equation}
q= -\frac{T_{01}}{\sqrt{-g_{tt}~g_{rr}}}~,~~\Delta=\sqrt{(\rho+p_{r}-2\eta\sigma_{11})^{2}-4q^{2}}.
\end{equation}
In addition, the weak, dominant and strong energy conditions (WEC, DEC and SEC) are to be satisfied, and these are given respectively as 
\begin{eqnarray}
(iv) &~&\rho - p_r + \Delta +2\eta\sigma_{11} \geq 0 ~~({\rm WEC}) \nonumber\\
(v) &~&\rho - p_r + 2\eta\sigma_{11} \geq 0 ~~({\rm DEC1}) \nonumber\\
(vi) &~&\rho - p_r -2p_{\theta} + \Delta+2\eta\left(\sigma_{11} + 2\sigma_{22}\right)\geq 0~~ ({\rm DEC2}) \nonumber\\
(vii) &~&\rho - p_r -2p_{\theta} + \Delta+2\eta\left(\sigma_{11} + 2\sigma_{33}\right)\geq 0 ~~({\rm DEC3}) \nonumber\\
(viii) &~&2p_{\theta} + \Delta - 2\eta\left(\sigma_{22} + \sigma_{33}\right) \geq 0 ~~({\rm SEC})
\label{energyconsgen}
\end{eqnarray}
where the DEC consists of three separate conditions labeled DEC1, DEC2 and DEC3. 
For convenience, we record the above conditions in the case of vanishing shear, and they read,
\begin{equation}
{\RN 1}.~|\rho +p_{r}|-2|q|\geq0~,~
{\RN 2}.~\rho-p_{r}+2p_{\theta}+\Delta\geq 0.
\label{con12}
\end{equation}
\begin{equation}
{\RN 3}.~\rho-p_{r}+\Delta\geq0~,~
{\RN 4}{\rm A}.~\rho-p_{r}\geq0~,~~{\RN 4}{\rm B}.~\rho-p_{r}-2p_{\theta}+\Delta\geq 0~,~
{\RN 5}.~~2p_{\theta}+\Delta\geq0.
\label{energycons}
\end{equation}

Finally, all the conditions above will need to be supplemented by the junction conditions in $f(R)$ models \cite{Deruelle},\cite{Senovilla}.
Recall that in GR, the standard Darmois-Israel junction conditions \cite{Israel} are valid, which amount to matching of the first
and second fundamental forms at the time-like hypersurface $\Sigma : r = r_0$. 
These are defined, with $a,b$ denoting the indices on the hypersurface, as, 
\begin{equation}
g_{ab}=g_{\alpha\beta}e^{\alpha}_{a}e^{\beta}_{b}~,~~
K_{ab}=\frac{1}{2}\mathcal{L}_{N}g_{ab}=\frac{1}{2}\left(g_{ab,c}N^{c}+g_{cb}N^{c}_{,a}+g_{ac}N^{c}_{,b}\right)~,
\end{equation}
where $N^{\mu}$ is the unit normal across the matching hypersurface,
Here, $e^{\alpha}_{a}=\frac{\partial x^{\alpha}}{\partial y^{a}}$ are tangents to the matching hypersurface, and 
$\mathcal{L}_{n}g_{ab}$ is the Lie derivative of the induced metric with respect to the normal vector to the hypersurface. 
For $f(R)$ collapse, the additional requirements are the continuity of the Ricci scalar and its first derivative across
this hypersurface \cite{Deruelle},\cite{Senovilla}, so that the full set of matching conditions across the collapsing
time-like hypersurface separating $ds^{2}_{-}$ and $ds^{2}_{+}$ are 
\begin{equation}
\left[g_{ab}\right]=0~,~\left[K_{ab}\right]=0~,~[R]=0~,~N^{\mu}[\partial_{\mu} R] = 0~,
\label{allmatch}
\end{equation}
where $[a]$ denotes the difference in the quantity $a$ across the hypersurface $\Sigma$. 
Therefore, in studying any model of $f(R)$ collapse, we will need to impose the junction conditions of 
eq.(\ref{allmatch}), in addition to the energy conditions spelt out in eqs.(\ref{con12gen}) and (\ref{energyconsgen}). 

The first two relations of eq.(\ref{allmatch}) are fairly straightforward to deal with. Since the analysis is standard, 
we will not go into the details here, but simply state the these imply that the Misner-Sharp mass function \cite{MS1},\cite{MS2} 
given by
\begin{equation}
M(r,t)=\frac{Q}{2}\left[1-e^{-2\psi}\left(\frac{dQ}{dr}\right)^{2}+e^{-2\nu}\left(\frac{dQ}{dt}\right)^{2}\right]~,
\end{equation}
equals the Schwarzschild mass when evaluated at the boundary $\Sigma$. 
From a fairly straightforward analysis, it is known that these also imply, from the metrics of eqs.(\ref{inner}) 
and (\ref{outer}) that 
\begin{eqnarray}
\frac{Q}{2}e^{-\left(\nu+\psi\right)}\left[2\frac{{\dot Q}'}{Q} -2 \frac{{\dot Q}}{Q}\frac{{\dot\psi}}{\psi} -2\frac{\nu'}{\nu}\frac{{\dot Q}}{Q}
+ e^{\left(\psi - \nu\right)}\left(2\frac{{\ddot Q}}{Q} - 2\frac{{\dot Q}}{Q}\frac{{\dot \nu}}{\nu} + \frac{e^{2\nu}}{Q^2} + \frac{{\dot Q}^2}{Q^2}
- e^{2\left(\nu - \psi\right)}\left(\frac{Q'^2}{Q^2} - 2\frac{\nu'}{\nu}\frac{Q'}{Q}\right)\right)\right]\biggr|_{\Sigma}=0
\end{eqnarray}
Equivalently, the junction conditions imply that \cite{Mars}
\begin{equation}
N^{\mu}\left[T_{\mu \nu}\right] = 0~,
\label{normal}
\end{equation}
which is a familiar condition in GR. The other two relations of eq.(\ref{allmatch}) are the essential new ingredients in this analysis. 
In summary, our task is to study collapse in $f(R)$ gravity, that are restricted by eight conditions mentioned in 
eqs.(\ref{con12gen}) and (\ref{energyconsgen}) in addition to the four junction conditions spelt out in eq.(\ref{allmatch}). Indeed, this
seems to be a formidable task, especially in cases with shear, but as we elaborate upon below, some simple solutions can 
nonetheless be found by utilising the constraints of eq.(\ref{allmatch}). 

\section{Separable interior solutions}

The extra junction conditions in $f(R)$ gravity are in fact quite strong, and can potentially exclude several well known 
collapse solutions in GR. For example, the Oppenheimer-Snyder solution is not an admissible collapsing solution
in modified gravity scenarios \cite{Senovilla}. 
As another concrete example, suppose we assume that the interior
metric is of the Lemaitre-Tolman-Bondi (LTB) form \cite{LTB1},\cite{LTB2}, \cite{LTB3} given by 
\begin{equation}
ds_-^2 = -dt^2 + X^2\left(r,t\right)dr^2 + Q^2\left(r,t\right)d\Omega^2
\label{LTB}
\end{equation}
where $X(r,t)$ and $Q(r,t)$ are functions of the co-moving radial coordinate and time, and $d\Omega^2$ is the
metric on the unit two-sphere. The special case of the 
homogeneous Friedmann-Robertson-Walker metric is obtained from eq.(\ref{LTB}) by writing 
\begin{equation}
X(r,t) = \frac{a(t)}{\sqrt{1-kr^2}}~,~Q(r,t) = a(t) r
\label{sep}
\end{equation}
with $k$ being a suitable constant. 
Also, the Einstein equations of GR can be shown to imply, for the general metric of eq.(\ref{LTB}), 
\begin{equation}
X(r,t) = {\mathcal A}(r)Q'(r,t)~,
\label{nonsep}
\end{equation}
with ${\mathcal A}(r)$ being an arbitrary function of the co-moving radial coordinate.\footnote{Here and otherwise, a prime 
will refer to a derivative with respect to the radial coordinate.} We will consider these two cases separately. 

We first consider a separable solution for the interior metric, of the form given in eq.(\ref{sep}), 
and assume that in co-moving coordinates, this is 
\begin{equation}
ds_-^{2}=-dt^2 + \frac{a(t)^{2}}{h(r)}dr^2 + a(t)^{2}r^{2}dr^2 + a(t)^{2}r^{2}\sin^2\theta d\phi^2~,
\label{metric}
\end{equation}
Since we are in co-moving coordinates we choose $u^{\mu}=(1,0,0,0)$ and $n^{\mu}=(0,\sqrt{h(r)}/a(t),0,0)$, so that the heat flux is 
along the radial direction, i.e $q^{\mu}=qn^{\mu}$.
With this metric, for the model described the the Lagrangian density $R+f(R)=R+\alpha R^{2}$ 
(where $\alpha$ is a constant) we can write down the energy momentum components:
\begin{eqnarray}\label{t00}
&~& \rho=(1+F)G_{00}-\frac{\alpha R^{2}}{2}-\frac{1}{2ra^{2}}\bigg(2rhF^{\prime\prime}+(4h+rh^{\prime}F^{\prime})
-6ra\dot{a}\dot{F}\bigg)~,
\nonumber\\
&~& \frac{a^{2}}{h}p_{r}=(1+F)G_{11}+\frac{a^{2}}{h}\frac{\alpha R^{2}}{2}+\frac{2F^{\prime}}{r}-\frac{a}{h}\bigg(a\dot{a}
\dot{F}+a\ddot{F}\bigg)~,
\nonumber\\
&~&a^{2}r^{2}p_{\theta}=(1+F)G_{22}+\frac{\alpha R^{2}}{2}a^{2}r^{2}-\frac{r}{2}\bigg(4a\dot{a}\dot{F}+2ra^{2}\ddot{F}-(2h+rh^{\prime})F^{\prime}-2rhF^{\prime\prime}\bigg)~,
~~ \frac{a}{\sqrt{h}}q=\dot{F}^{\prime}-\frac{\dot{a}}{a}F^{\prime}.
\end{eqnarray}
The Ricci scalar of the interior metric is calculated to be
\begin{equation}
R(r,t)=\frac{2}{a^{2}}\left(\frac{1-h-rh^{\prime}}{r^{2}}\right)+6\frac{\left(\Dot{a}^{2}+a\Ddot{a}\right)}{a^{2}}.
\label{Ricci}
\end{equation}
In order that the Ricci scalar matches smoothly to the collapsing co-moving boundary at all co-moving times, 
we will therefore require that $\Dot{a}^{2}+a\Ddot{a} = 0$ (since the second term on the right hand side of eq.(\ref{Ricci})
is a function of time only), in which case the first term of eq.(\ref{Ricci})
can be appropriately solved in order to fulfil the requirement that $R$ is continuous across the matching hypersurface.
However, to satisfy eq.(\ref{normal}), one finds after a straightforward calculation, using the unit normal
vector $N^{\mu} = \left(0, \sqrt{h(r)}/a(t),0,0\right)$, the condition 
\begin{equation}
h(r) -1-r^2\left(\Dot{a}^{2}+2a\Ddot{a}\right)=0~.
\end{equation}
In order to satisfy this for all times, one thus requires $\Dot{a}^{2}+2a\Ddot{a} = 0$ which naturally implies that
this cannot be satisfied in conjunction with the criterion for a continuous Ricci scalar across the boundary, at all co-moving times. 
In conclusion, what we have here is a no go scenario, namely that a simple separable form of the metric given in eq.(\ref{metric}) 
is unsuitable for describing collapse in $f(R)$ gravity. 

The assumption of a separable solution of the form in eq.(\ref{metric}) is possibly an over-simplification. We will next
consider another separable form of the interior metric given by 
\begin{equation}
ds_-^2 = -A(r)^2dt^2 + 2a(t)^2\left(\partial_r A(r)\right)^2dr^2 + a(t)^2A(r)^2d\Omega^2~,
\label{Wagh}
\end{equation}
with the energy momentum tensor having the same form as in eq.(\ref{emt}). We will match this with an external
Schwarzschild solution. 
This metric was originally considered in \cite{WaghMaharaj} to the study of the collapse of a shear-free radiating
spherically symmetric star in GR. As we elaborate below, this ansatz offers considerable simplifications in the study of
collapsing stars in $f(R)$ gravity. 

To this end, we first note that the Ricci scalar here is given by 
\begin{equation}
R = -\frac{1-6\left({\Dot a(t)}^2 + a(t){\Ddot a(t)}\right)}{a(t)^2A(r)^2}
\label{Riccisep}
\end{equation}
Using eqs.(\ref{Eeq}) and (\ref{emt}), the relevant physical quantities are obtained for the metric of eq.(\ref{Wagh}) as
\begin{eqnarray}
A(r)^{2} \rho&=&\big(1+F\big)\bigg(\frac{1+6\dot{a}^{2}}{2a^{2}}\bigg)+\frac{\alpha A^{2} R^{2}}{2}-\bigg(\frac{A\big((2A^{\prime2}-AA^{\prime\prime})F^{\prime}+AA^{\prime}F^{\prime\prime}\big)-6a\dot{a}A^{\prime3}\dot{F}}{2a^{2}A^{\prime3}}\bigg)~,~\nonumber\\
2a^{2}A^{\prime 2} p_{r}&=&\big(1+F\big)\bigg(\frac{A^{\prime2}(1-2\dot{a}^{2}-4a\ddot{a})}{A^{2}}\bigg)+\alpha a^{2}A^{\prime 2}R^{2}-\bigg(\frac{A^{\prime}\big(2aA^{\prime}(2\dot{a}\dot{F}+a\ddot{F})-3AF^{\prime}\big)}{A^{2}}\bigg)~,~\nonumber\\
a^{2}A^{2}p_{\theta}&=&\big(1+F\big)\bigg(\frac{1-2\dot{a}^{2}-4a\ddot{a}}{2}\bigg)+\frac{\alpha R^{2}a^{2}A^{2}}{2}-a(2\dot{a}\dot{F}
+ a\ddot{F})-\bigg(\frac{A\big((AA^{\prime\prime}-2A^{\prime2})F^{\prime}-AA^{\prime}F^{\prime\prime}\big)}{2A^{\prime 3}}\bigg)~,~
\nonumber\\
 -\sqrt{2}aAAq&=&\big(1+F\big)\frac{2\dot{a}A^\prime{}}{aA}+\frac{A^{\prime}\dot{F}}{A}+\frac{\dot{a}F^{\prime}}{a}-\dot{F}^{\prime}~.
\label{allterms}
\end{eqnarray}
Now, from the metric of eq.(\ref{Wagh}), it can be checked via eq.(\ref{allterms})
that the pressure anisotropy is given by
\begin{equation}
p_{\theta} - p_r = -\frac{8\alpha\left[1-6\left({\Dot a(t)}^2 + a(t){\Ddot a(t)}\right)\right]}{a(t)^4A(r)^4}~.
\label{anisosep}
\end{equation}
Importantly, if we demand that the pressure anisotropy vanishes identically, then it necessarily implies that $R=0$, in which case
the solution reduces to one in GR. We are therefore naturally constrained to consider situations with pressure
anisotropy in $f(R)$ scenarios.

It is then seen that in order for the Ricci scalar and its derivative to be continuous across the matching hypersurface
(which we choose without loss of generality to be $r = 1$), it is enough for us to choose $A(r) = (1-r)^{-n}$ with $n\geq 1$,
so that continuity of the Ricci scalar and its derivative is guaranteed at the boundary. 
The function $a(t)$ is unspecified at this stage. In order to simplify the computations, we will have to make a choice, 
and to this end we will choose ${\Dot a(t)}^2 + a(t){\Ddot a(t)}=0$. To summarise, our ansatz for a solution of
the metric of eq.(\ref{Wagh}) is (with $b$ and $n$ being constants),
\begin{equation}
a(t) = \sqrt{1-2bt}~,~~A(r) = \frac{1}{(1 - r)^n}~,~~n \geq 1 ~,~~b>0
\label{ansatz1}~.
\end{equation}
We will henceforth choose for simplicity, the constants
$b = 1/2$ and $n=2$ so that $R=-\left(1-r\right)^4/\left(1-t\right)$ and satisfies both the conditions on the Ricci scalar
mentioned in eq.(\ref{allmatch}) at the boundary, arbitrarily close to the time of collapse. In this notation, the collapse starts
at $t=0$ and a singularity forms at $t=1$ where the scale factor $a(t)$ goes to zero and the Ricci scalar diverges 
although all co-moving observers see an apparent horizon at $t=1/2$, as we elaborate in a while. As usual, our
interior solution is matched to an external Schwarzschild metric at $r=1$.

Now, using the fact that the hypersurface normal is given by the vector
\begin{equation}
N^{\mu} = \left(0,\frac{1}{\sqrt{2} a(t)A'(r)},0,0\right)~,
\end{equation}
it can be immediately seen that the condition $N^{\mu}\left[T_{\mu\nu}\right]=0$ is satisfied at all times. 
Now upon using eqs.(\ref{Riccisep}) and (\ref{ansatz1}), we finally obtain the very simple expressions,
\begin{eqnarray}
\rho &=& \frac{\left(1-r\right)^4 \left(5-7t+2t^2 +2 \alpha  (1-r)^4 (4-t)\right)}{4 \left(1-t\right)^3}~,~~
p_r =\frac{(1-r)^4 \left(3-5t + 2t^2 + 2 \alpha(1-r)^4 (14-11t)\right)}{4 \left(1-t\right)^3}~,~\nonumber\\
p_{\theta} &=& \frac{(1-r)^4 \left(3-5t + 2t^2 - 2 \alpha(1-r)^4 (2-5t)\right)}{4 \left(1-t\right)^3}~,~~
q=-\frac{(1-r)^4 \left(t-1-6 \alpha  (1-r)^4\right)}{\sqrt{2} (1-t)^{5/2}}~.
\label{qtys}
\end{eqnarray}
It is clearly seen from eq.(\ref{qtys}) that all the components of the stress tensor vanish at the boundary $r=1$, and that
the pressure and density are positive for all values of the co-moving radius, at all co-moving times.\footnote{These 
diverge at the time of formation of the singularity, as expected} This situation
thus corresponds to the realistic collapse of a dense star in $f(R)$ gravity. 

This last statement requires some clarification. From our discussion above, it follows that the collapse reaches 
a singularity in co-moving time $t=1$, when $a(t)=0$ and the Ricci scalar diverges at all co-moving radii. This is a 
shell focusing singularity, which happens simultaneously for all co-moving observers. In order to determine whether the 
singularity is naked or not, we have to investigate the formation of trapped surfaces during the collapse process. 
These are the compact two-dimensional space-like surfaces such that both families of ingoing and outgoing null 
geodesics orthogonal to them necessarily converge. Mathematically one can find out such locations from the expansion 
parameter $\Theta$ of the outgoing future-directed null geodesics. We consider a congruence of 
outgoing radial null geodesics having the tangent vector $\left(u^t, u^r, 0, 0\right)$. If such geodesics terminate at the 
singularity in the past with a definite tangent vector, then at singularity we have $\Theta > 0$. 
When such curves do not exist it means that an event horizon has formed earlier than singularity, thus forming a 
blackhole as the end stage of the collapse process. 

Now recall that for a spherically symmetric metric such as the one we are considering, the co-moving time of 
formation of an apparent horizon is given from the equation
\begin{equation}
g^{\mu\nu}\partial_{\mu}Q(r,t)\partial_{\nu}Q(r,t) =0
\end{equation}
Using eq.(\ref{Wagh}) and the ansatz of eq.(\ref{ansatz1}), it is then checked that the co-moving time formation
of the apparent horizon for all co-moving observers is $t=1/2$. Hence, the end state of the collapse process is 
a black hole in this case. 
%%%%%%%%%%%%%%%%%%%%%%%%%%%%%%%%%%%%%%%%%%%%%%%%%%%%%%%%%%%%%
\begin{figure}[t!]
\begin{minipage}[b]{0.3\linewidth}
\centering
\includegraphics[width=2in,height=1.6in]{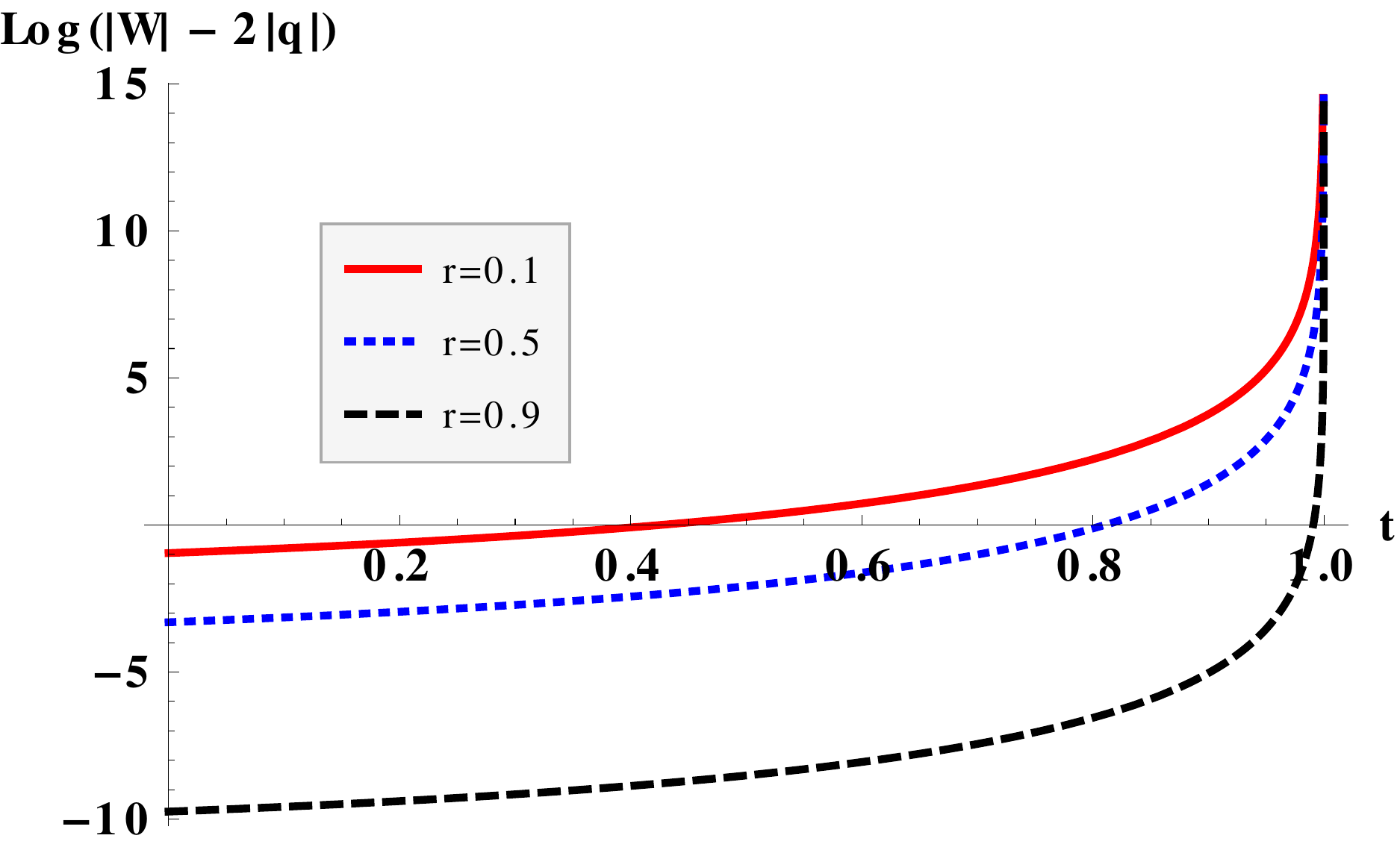}
\caption{Condition {\RN 1}}
\label{con1}
\end{minipage}
\hspace{0.2cm}
\begin{minipage}[b]{0.3\linewidth}
\centering
\includegraphics[width=2in,height=1.6in]{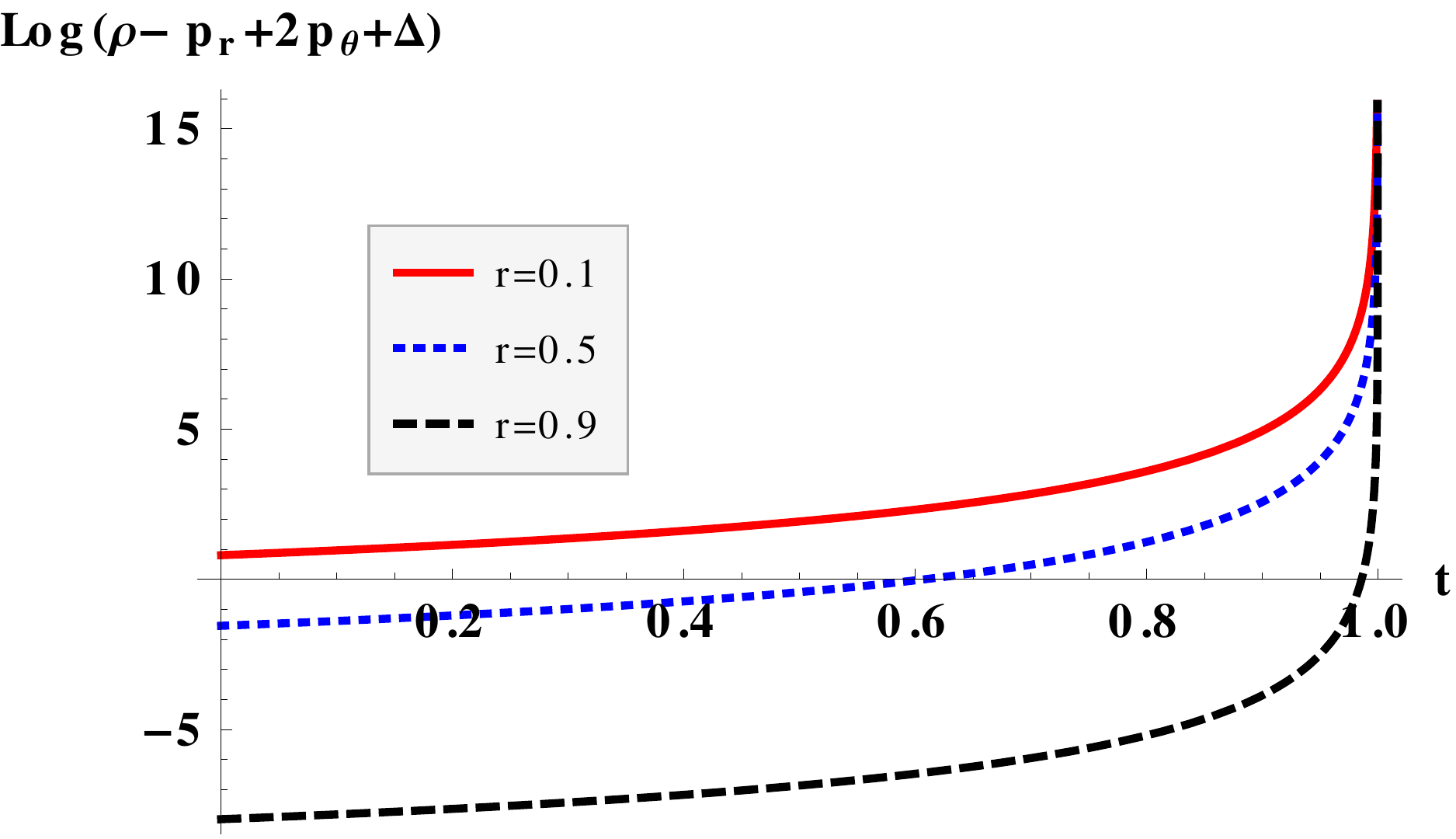}
\caption{Condition {\RN 2}}
\label{con2}
\end{minipage}
\hspace{0.2cm}
\begin{minipage}[b]{0.3\linewidth}
\centering
\includegraphics[width=2in,height=1.6in]{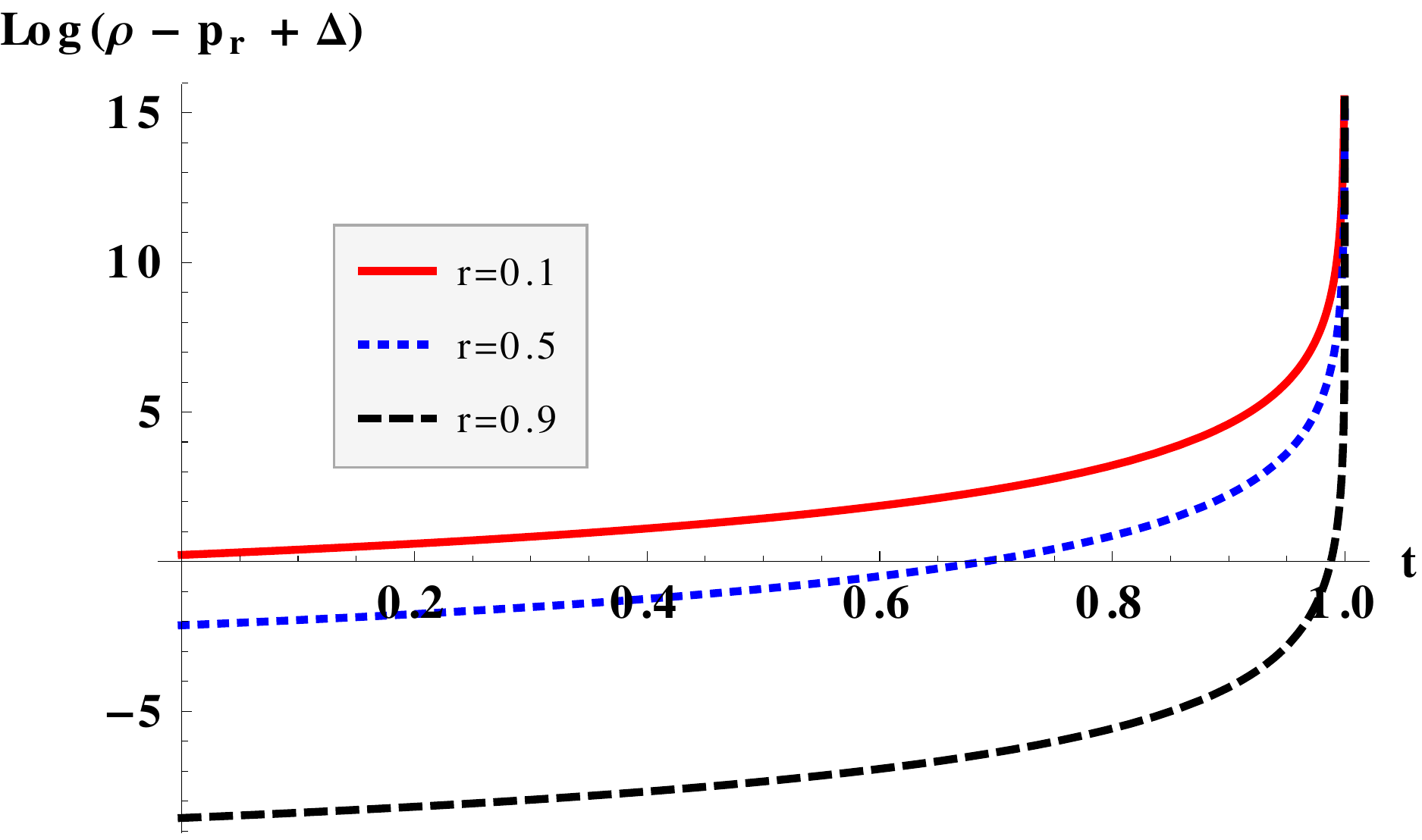}
\caption{Condition {\RN 3}}
\label{wec}
\end{minipage}
\end{figure}
%%%%%%%%%%%%%%%%%%%%%%%%%%%%%%%%%%%%%%%%%%%%%%%%%%%%%%%%%%%%%
%%%%%%%%%%%%%%%%%%%%%%%%%%%%%%%%%%%%%%%%%%%%%%%%%%%%%%%%%%%%%
\begin{figure}[t!]
\begin{minipage}[b]{0.3\linewidth}
\centering
\includegraphics[width=2in,height=1.6in]{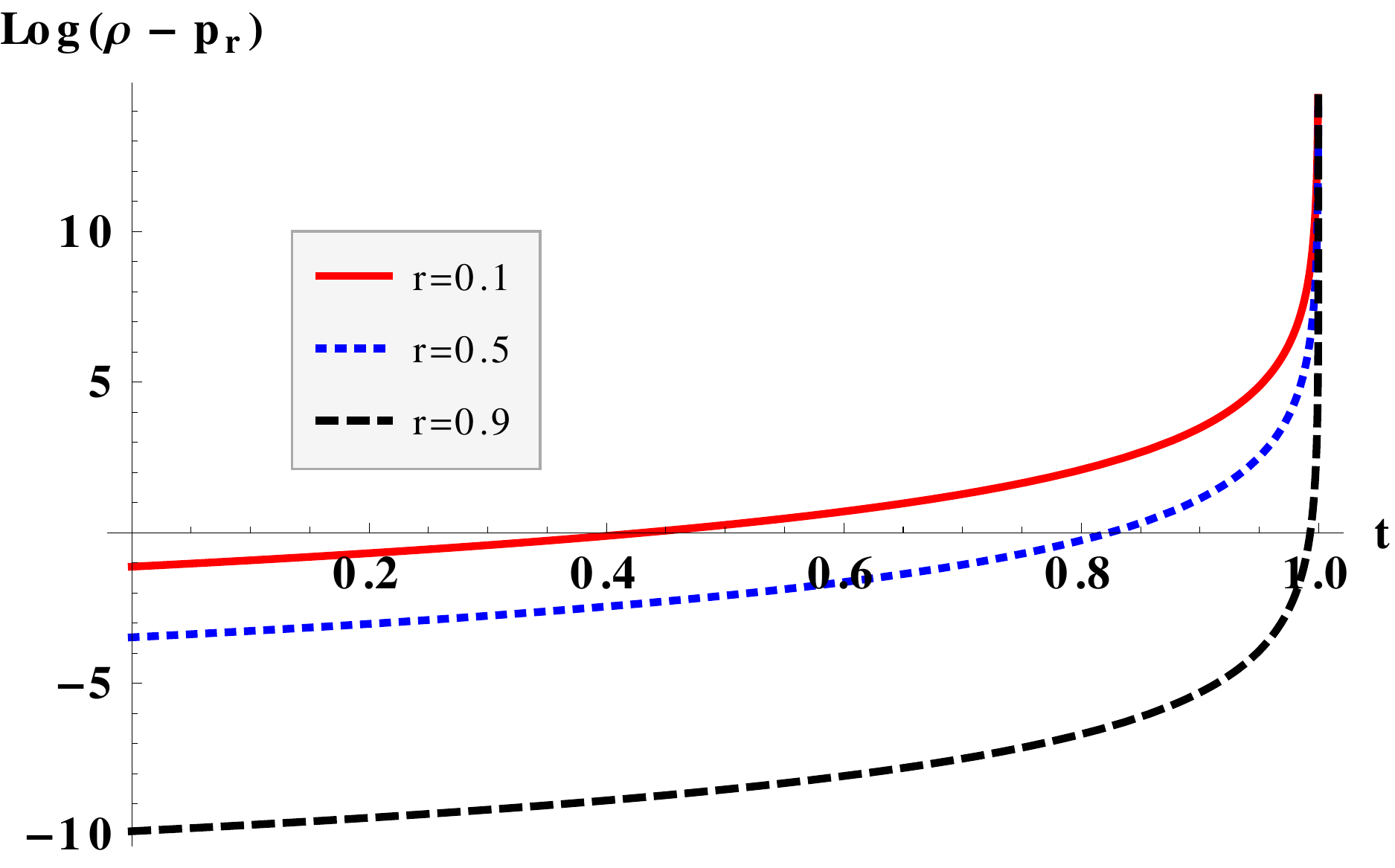}
\caption{Condition {\RN 4}A}
\label{dec1}
\end{minipage}
\hspace{0.2cm}
\begin{minipage}[b]{0.3\linewidth}
\centering
\includegraphics[width=2in,height=1.6in]{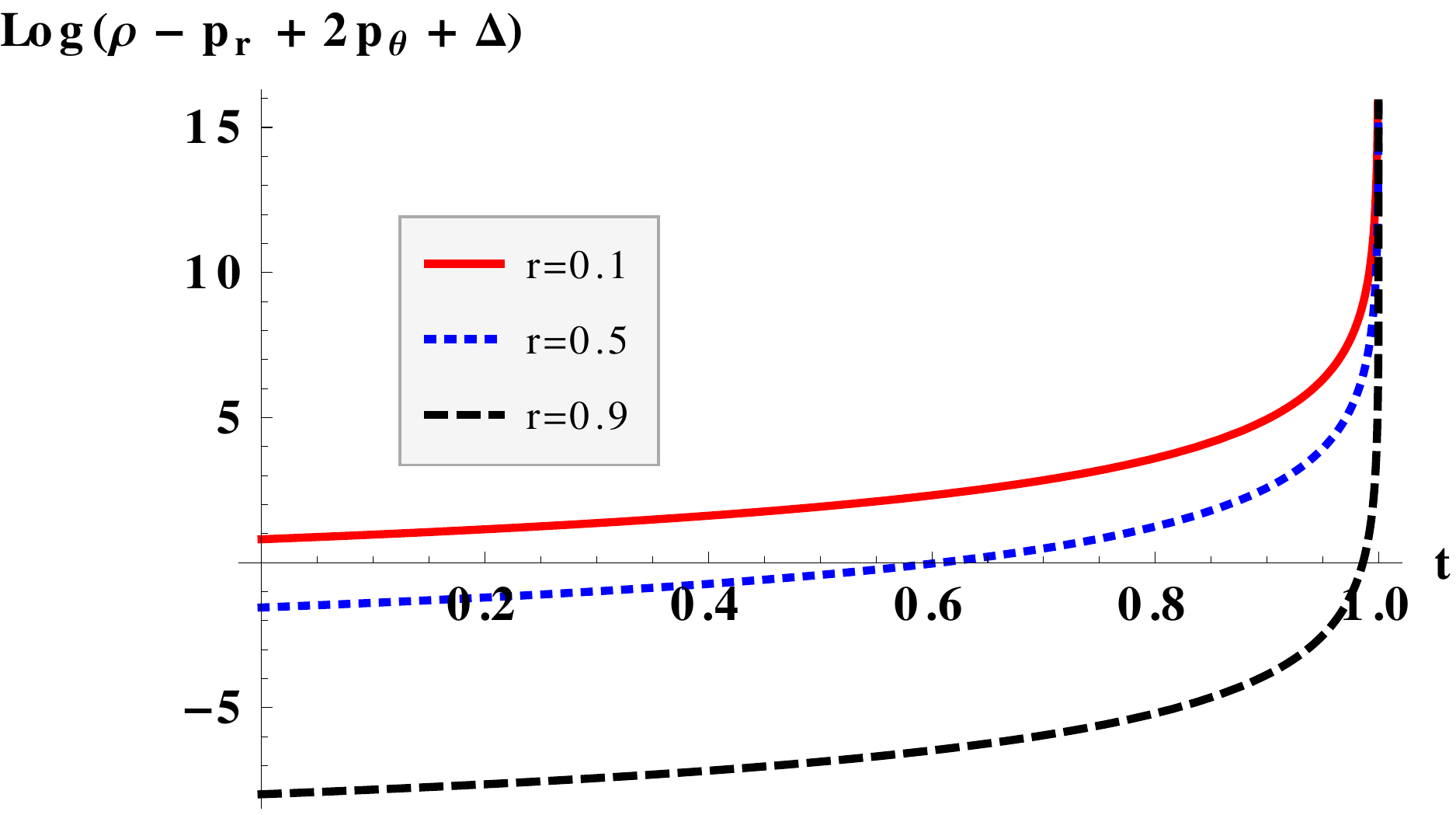}
\caption{Condition {\RN 4}B}
\label{dec2}
\end{minipage}
\hspace{0.2cm}
\begin{minipage}[b]{0.3\linewidth}
\centering
\includegraphics[width=2in,height=1.6in]{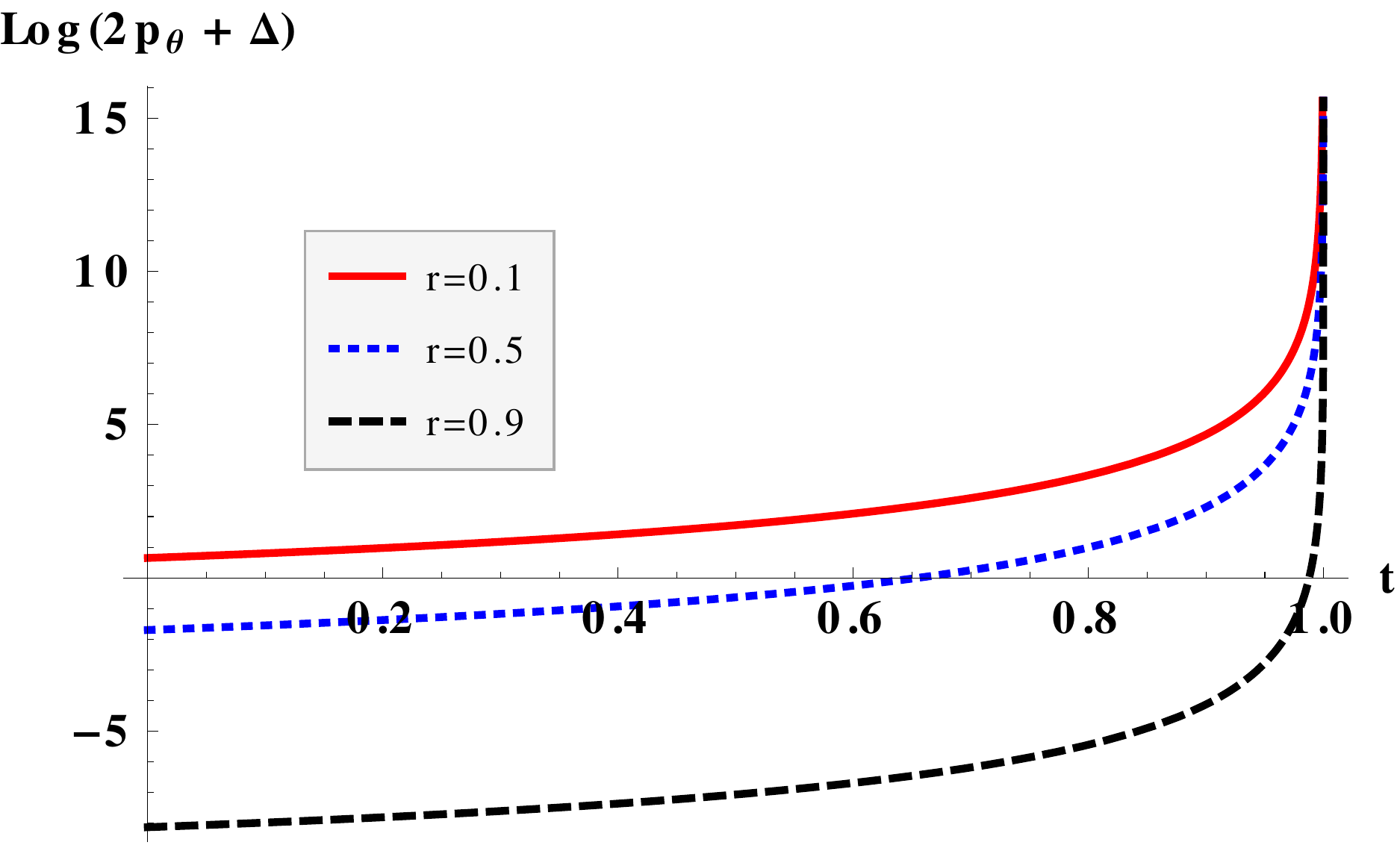}
\caption{Condition {\RN 5}}
\label{sec}
\end{minipage}
\end{figure}
%%%%%%%%%%%%%%%%%%%%%%%%%%%%%%%%%%%%%%%%%%%%%%%%%%%%%%%%%%%%
This is also obtained by computing the boundary redshift for an observer at infinity, which diverges at the formation
time of the black hole. This is obtained by writing the external Schwarzschild solution of eq.(\ref{outersch}) in terms of 
retarded time and computing the junction conditions, and the co-moving time for our collapsing scenario at which the
redshift at infinity diverges is \cite{Chan0},\cite{Chan}
\begin{equation}
\frac{1}{\sqrt{2}} - \frac{1}{2\sqrt{1-t}} = 0~,~~
\end{equation}
which yields the same result $t=1/2$. 

It remains to check the validity of the energy conditions listed in eqs.(\ref{con12}) and (\ref{energycons}). This is
most conveniently done numerically, since analytical experssions for these conditions become cumbersome. 
In this analysis,we choose $\alpha = 10^{-3}$. 
In figs.(\ref{con1}) to (\ref{sec}), we show that all the energy conditions are indeed satisfied. In all these figures,
the solid red, dotted blue and dashed black curves indicate the co-moving observer at $r=0.1$, $0.5$ and $0.9$, respectively.
We have shown the validity of the energy conditions from $t=0$ the $t=1$, although it is to be noted that as we
have discussed, the apparent horizon forms at $t=1/2$ for this model. 

Here, the four-velocity, and the unit vector in the radial direction are
\begin{equation}
u^{\mu} = \left((1-r)^2,0,0,0\right)~,~~n^{\mu}=\left(0,\frac{(1-r)^3}{2\sqrt{2}\sqrt{1-t}},0,0\right)
\label{uandn}
\end{equation}
These will satisfy the conditions $u^{\mu}u_{\mu} = -1$ and $n^{\mu}n_{\mu}=1$, along with those mentioned after
eq.(\ref{emt}). 
The shear tensor is identically zero in this case, as is generally true for separable solutions of the form that we consider here. 
It is also straightforward to check that the expansion scalar for a time-like congruence is given by
\begin{equation}
\Theta = -\frac{3\left(1-r\right)^2}{2\left(1-t\right)}~,
\end{equation}
Also, using eq.(\ref{uandn}), it can be checked that for our metric of eq.(\ref{Wagh}), the condition of eq.(\ref{finalmod}) is satisfied with 
\begin{equation}
{\mathcal E} = -\frac{\left(1-r\right)^4}{\left(1-t\right)}~,~~
\Pi = -\frac{8\alpha\left(1-r\right)^8}{\left(1-t\right)^2}~.
\label{Wagheq}
\end{equation}
With these inputs, it can be checked that eq.(\ref{finalmod1}) is indeed satisfied in this case, with $\sigma = 0$, and
so is eq.(\ref{finalmod2}). 

Note that here the pressure anisotropy goes to zero at the matching hypersurface as it should, but does not vanish at the origin ($r=0$). 
Interestingly, this is an artefact of $f(R)$ gravity, as the anisotropy vanishes identically with $\alpha = 0$, as follows
from eq.(\ref{anisosep}) or (\ref{Wagheq}). In this context, we note that
anisotropy in static situations (for example in compact stars) have been studied extensively (see, e.g. \cite{Her2},\cite{Her3}). It is
well known that in such static situations, the pressure anisotropy must vanish at the center, and that a non-zero central anisotropy
implies that the density at the center vanishes \cite{Madsen}. These conditions need not be satisfied in non-equilibrium situations
that we are considering here.
In this context, observe from eq.(\ref{finalmod2}) that since the shear is identically zero in this case, the anisotropy at the 
center is forced to be non-zero, since none of the terms in that equation vanish identically at $r=0$. This seems to be a generic
feature of $f(R)$ collapse.  

\section{Non-separable interior solutions}

We will now consider matching of Ricci scalar and its derivatives with a non-separable spherically symmetric  
metric of the form given in eq.(\ref{nonsep}). For convenience, we write ${\mathcal A}(r) = (1-h(r))^{-1}$, and thus we have
our ansatz for the interior metric 
\begin{equation}
ds^{2}=-dt^{2}+\frac{Q^{\prime2}}{1-h(r)}+Q^{2}(r,t)d\Omega^{2}~,
\end{equation}
where $Q(r,t)$ is the co-moving radius of the collapsing matter, and $h(r)$ is function of $r$ only. This metric has 
the form of a general LTB solution.
We can calculate the Ricci scalar as
\begin{equation}
R(r,t)=\frac{1}{Q^{2}Q^{\prime}}\frac{d}{dr}\left[2Q\left(\Dot{Q}^{2}+Q\Ddot{Q}+h(r)\right)\right]~.
\end{equation}
Since we want to match Ricci scalar and it's derivative across a junction, as the simplest choice, we put
\begin{equation}\label{qrt1}
\Dot{Q}^{2}+Q\Ddot{Q}=0~.
\end{equation}
Then, the Ricci scalar takes the simple form
\begin{equation}
R(r,t)=\frac{1}{Q^{2}Q^{\prime}}\frac{d}{dr}\left(2Qh\right)~.
\label{Riccisimple}
\end{equation}
The solution of eq.(\ref{qrt1}) is 
\begin{equation}
Q(r, t)= r\sqrt{g(r)-2bf(r)\left(t-t_{0}\right)}~,
\label{qrtnonsep}
\end{equation}
where $b >0$ is a constant and $f(r)$ and $g(r)$ are two (positive) function of $r$, which we have to choose. Without loss of generality,
we will henceforth set $b=1/2$, along with $t_0=0$, so that our collapse process begins at the origin of the co-moving time. 
	
Also we need to take $h(r)$ in such a way  that both Ricci scalar and it's derivative are continuous across the 
junction at $r_{0}$. We will make a simple choice here, and set
\begin{equation}
h(r)=(r_{0}-r)^{2}~,~~g(r) = \left(r_0-r\right)^{-4}~,~~f(r) = \left(r_0-r\right)^{-2}~.
\end{equation}
With this choice, from eq.(\ref{Riccisimple}), the Ricci scalar reads,
\begin{equation}
R = \frac{2\left(r_0-r\right)^7\left[1-\left(2r^2 - 3rr_0 + r_0^2\right)t\right]}{r^2\left[ r+r_0 - \left(r-r_0\right)^2\left(r + 2r_0\right)t
+ \left(r - r_0\right)^4r_0t^2\right]}~.
\end{equation}
It is then seen that continuity of the Ricci scalar and its derivative is guaranteed across the co-moving boundary, which for
simplicity we will now choose as $r_0=1$. Note that at $t=0$, there is an initial singularity at the origin, and the Ricci scalar
diverges as $R \sim 1/r^2$. We will however concentrate on the singularity that forms due to the collapse process, in which
case $R \sim 1/r^3$ near the origin, at the time of formation of the central singularity. However, we note that the process 
described in this section may not correspond to the realistic collapse of a dense star, contrary to the analysis of the 
previous section. 

To this end, note that this singularity forms along the curve $t=t_{s}(r)$ defined by
\begin{equation}
Q(t_{s}(r), r)=0 ~~{\rm i.e.}~~ t_{s}(r)=\frac{1}{\left(1-r\right)^2}~,
\label{tsr}
\end{equation}
and the co-moving time for the formation of the apparent horizon $t_{ah}(r)$ is given by 
\begin{equation}
t_{ah}(r)=\frac{8-5r}{4\left(2-r\right)\left(1-r\right)^2}
\label{tahr}
\end{equation}
This implies that in the reference frame of a co-moving observer (at fixed $r$), 
the singularity formation is not simultaneous (note that it was simultaneous in the case of  separable solutions), 
rather it is a curve in the $t-r$ plane which starts at $(t,r)=(1,0) $.  
If the apparent horizon starts forming at a co-moving time that is earlier than that of singularity formation, then the event 
horizon can fully cover the singularity and the end stage is a  black hole. On the other hand, if trapped 
surfaces form after the singularity, then it is possible that a non-space-like geodesic might come out of the 
singularity to reach an external observer, and in that case the final singularity will be visible, i.e the fate of
the collapse will be at least a locally naked singularity. 
%%%%%%%%%%%%%%%%%%%%%%%%%%%%%%%%%%%%%%%%%%%%%%%%%%%%%%%%%%%%%%%%%
\begin{figure}[t!]
\begin{minipage}[b]{0.3\linewidth}
\centering
\includegraphics[width=2in,height=1.6in]{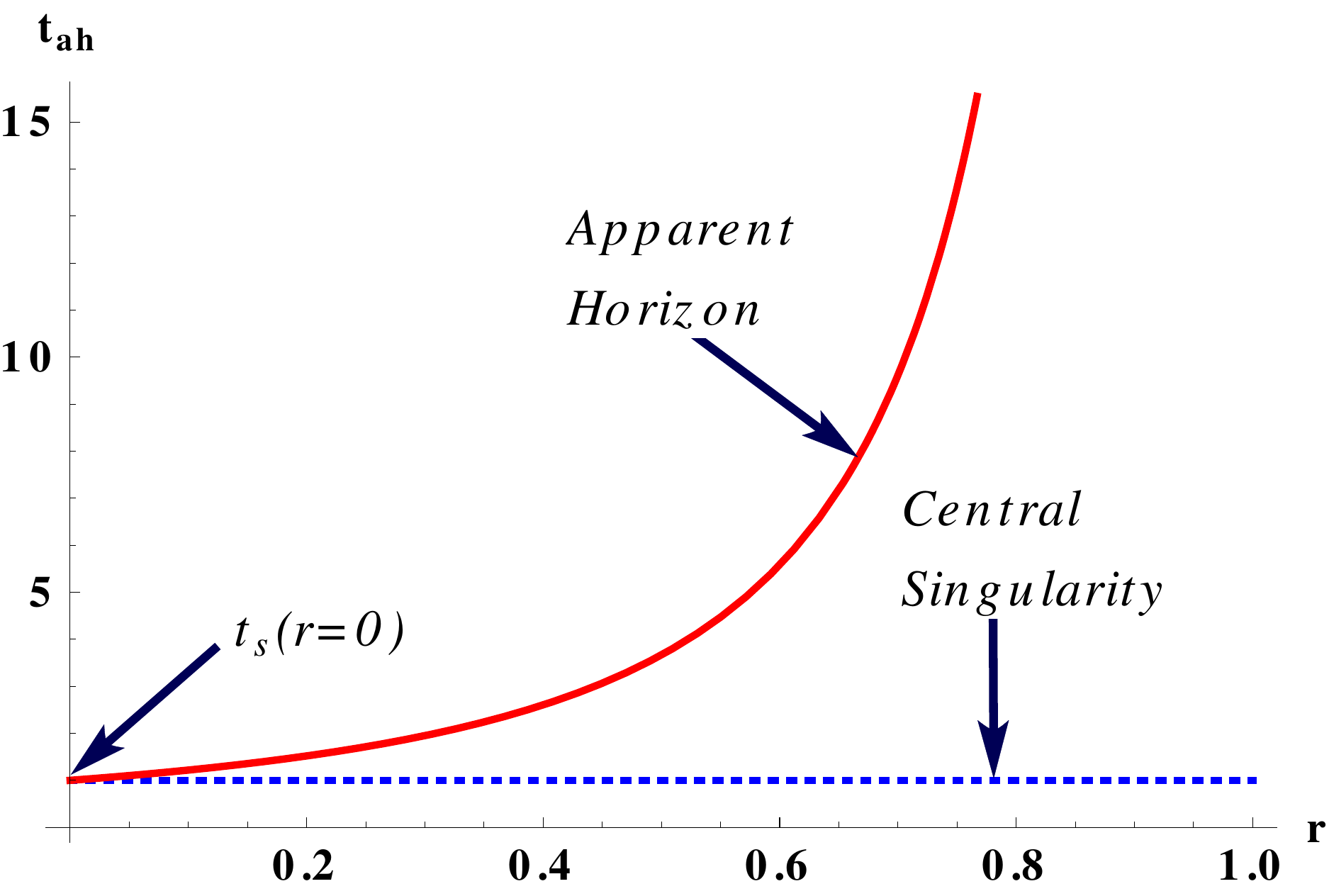}
\caption{$\rho$ vs $t$}
\label{tahns}
\end{minipage}
\begin{minipage}[b]{0.3\linewidth}
\centering
\includegraphics[width=2in,height=1.6in]{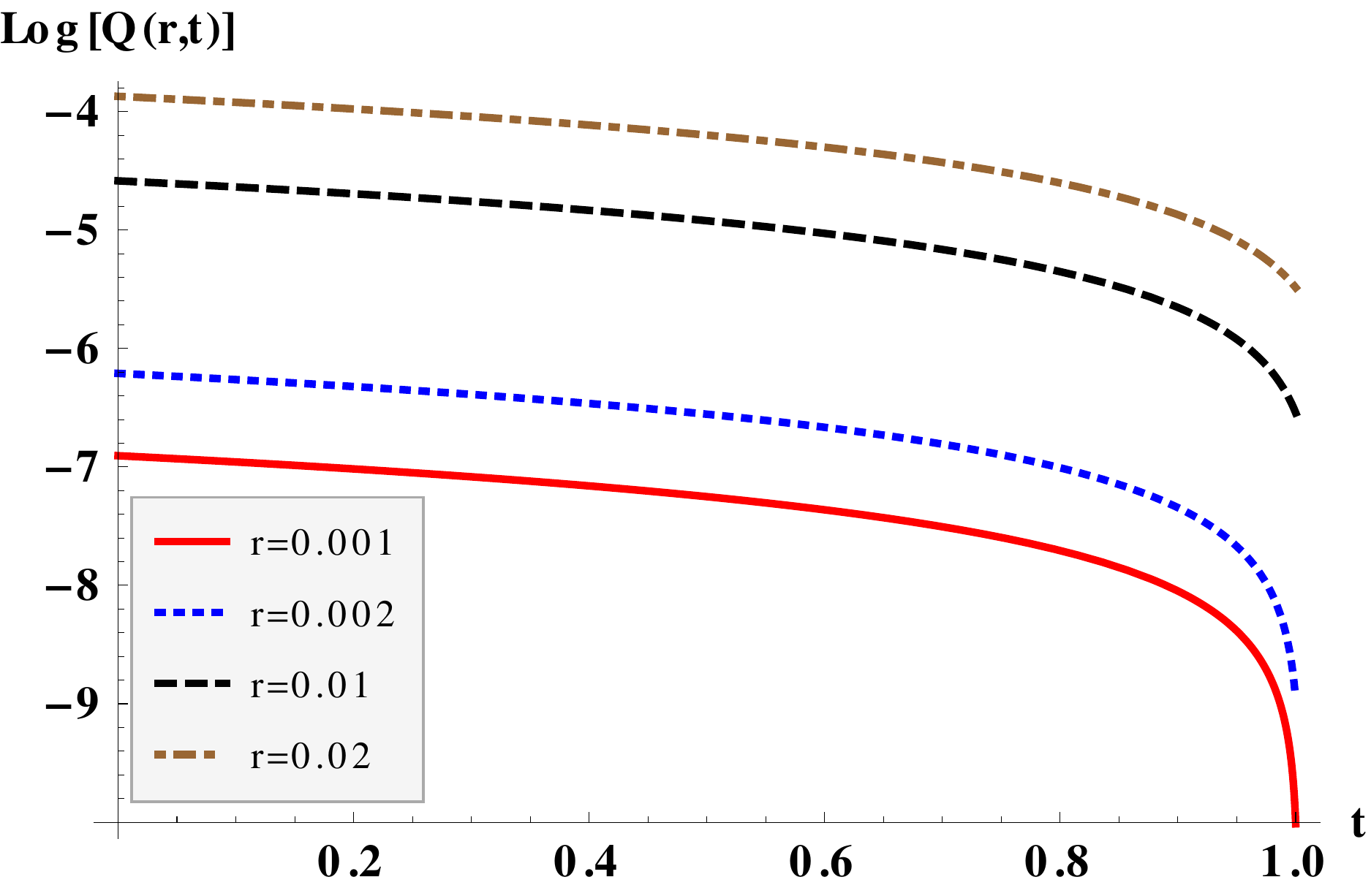}
\caption{$Q(r,t)$ vs t}
\label{qrtns}
\end{minipage}
\begin{minipage}[b]{0.3\linewidth}
\centering
\includegraphics[width=2in,height=1.6in]{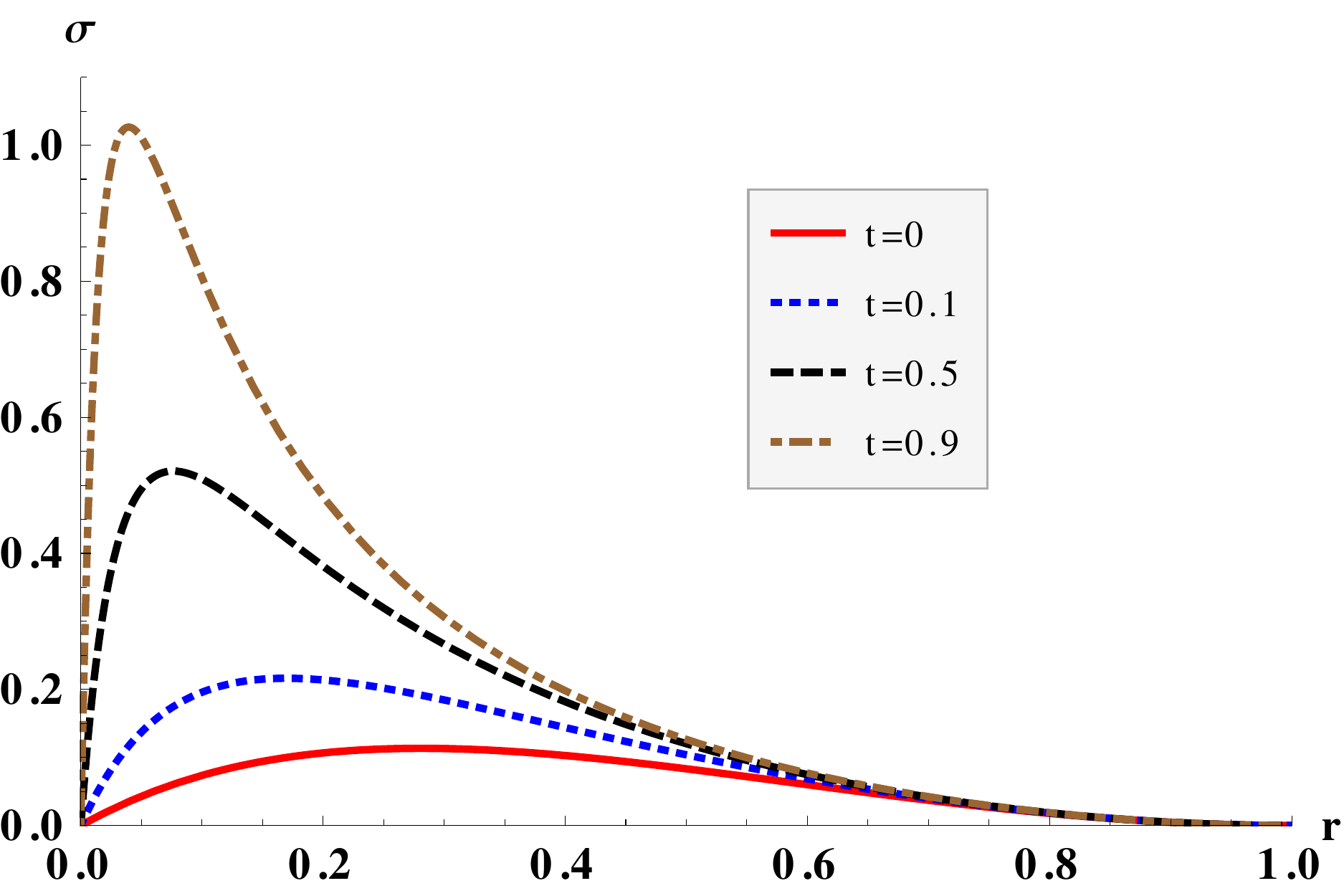}
\caption{$\sigma$ vs t}
\label{sigmans}
\end{minipage}
\end{figure}
%%%%%%%%%%%%%%%%%%%%%%%%%%%%%%%%%%%%%%%%%%%%%%%%%%%%%%%%%%%%%%

In fig.(\ref{tahns}), we show the apparent horizon curve of eq.(\ref{tahr}) as a function of time. This is shown in red, and the
dotted blue line is the time of formation of the central singularity, i.e $t_s(r=0)=1$. Clearly, all co-moving observers will 
see the formation of the central singularity first, and therefore conclude that the collapse results in a singularity
that is locally naked. In fig.(\ref{qrtns}),
we show the logarithm of $Q(r,t)$ as a function of time. Here, the thick red, dotted blue, dashed black and dot-dashed brown 
curves correspond to $r=0.001$, $0.002$, $0.01$ and $0.02$, respectively. 

The expansion scalar for a time-like geodesic congruence is calculated to be 
\begin{equation}
\Theta = \frac{3 (r-1)^4 t-(r-1)^2 (r+3)}{2 \left((r-1)^4 t^2-(r+2) (r-1)^2
t+r+1\right)}~,~~{\rm i.e}~~ \Theta\big|_{r \to 0} = -\frac{3}{2\left(1-t\right)}~,
\end{equation}
showing the central divergence at $t_s(r=0)=1$. Also, we record the expression for the shear,
\begin{equation}
\sigma=r\left(1-r\right)^2\left[1+r-t\left(1-r\right)^2\left(2+r\right)+ t^2\left(1-r\right)^4\right]^{-1}
\label{shear}
\end{equation}
In fig.(\ref{sigmans}), we show the behaviour of $\sigma$ as a function of $r$ for $t=0$ (thick red), $0.1$ (dotted blue), $0.5$ (dashed
black) and $0.9$ (dot dashed brown). Clearly, as the collapse progresses in co-moving time, the shear which was initially regular
at the center increases near the origin, and diverges as $\sigma \sim r^{-1}$ at the origin for $t \to 1$, as can be seen
from eq.(\ref{shear}). 
%%%%%%%%%%%%%%%%%%%%%%%%%%%%%%%%%%%%%%%%%%%%%%%%%%%%%%%%%%%%%%%%%
\begin{figure}[t!]
\begin{minipage}[b]{0.3\linewidth}
\centering
\includegraphics[width=2in,height=1.6in]{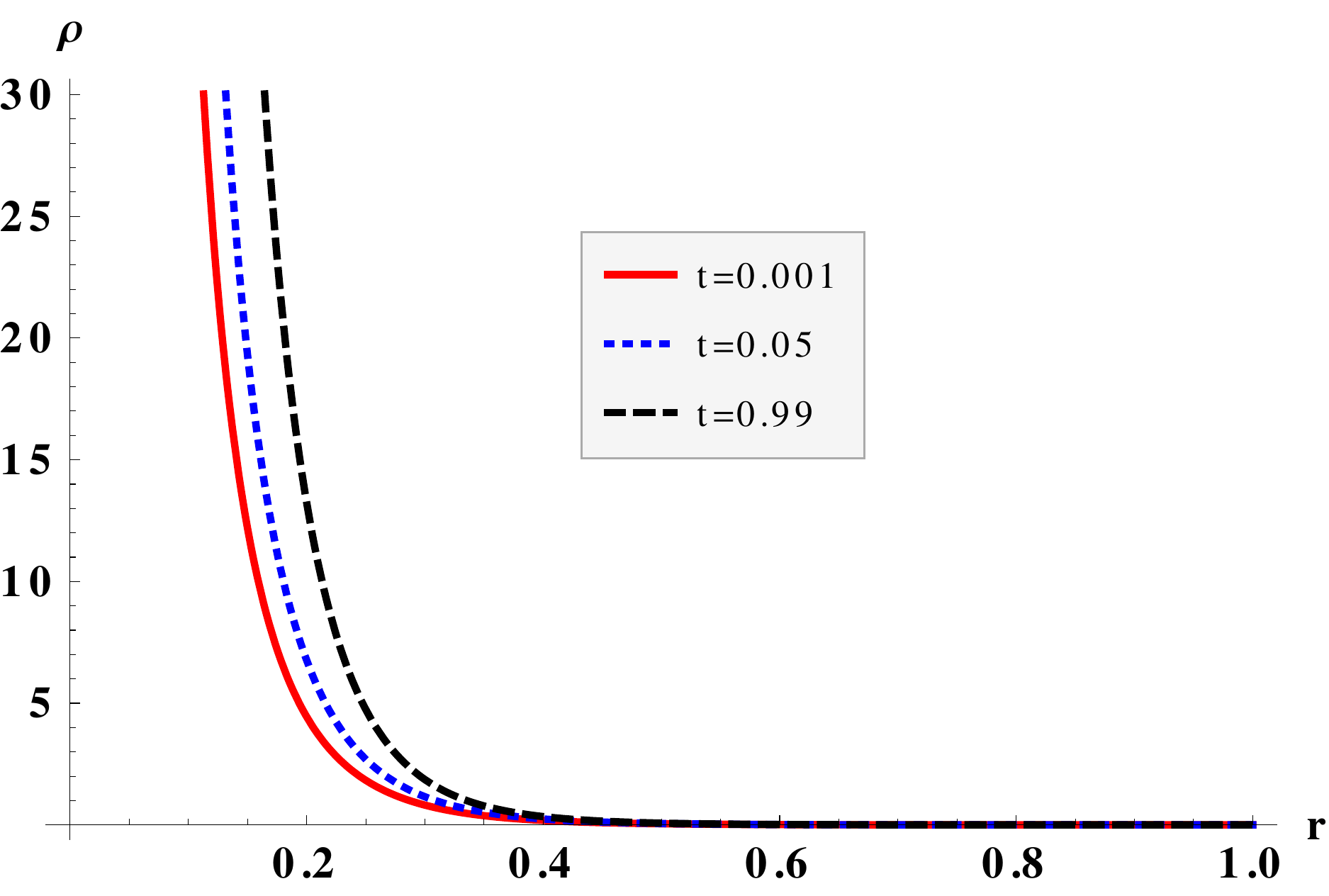}
\caption{$\rho$ vs $r$}
\label{rhons}
\end{minipage}
\hspace{0.2cm}
\begin{minipage}[b]{0.3\linewidth}
\centering
\includegraphics[width=2in,height=1.6in]{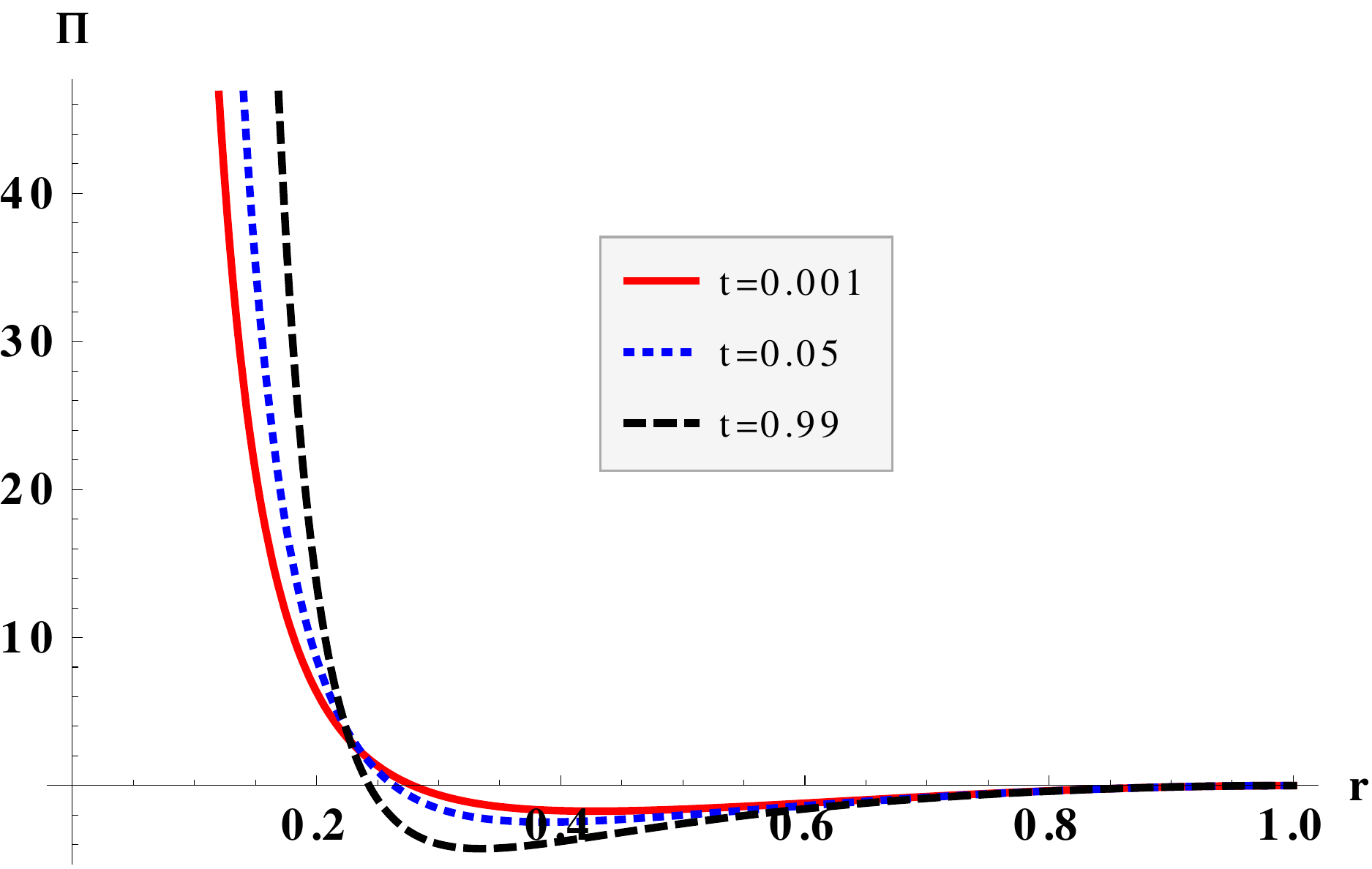}
\caption{$p_{\theta} - p_r$ vs r}
\label{anisons}
\end{minipage}
\hspace{0.2cm}
\begin{minipage}[b]{0.3\linewidth}
\centering
\includegraphics[width=2in,height=1.6in]{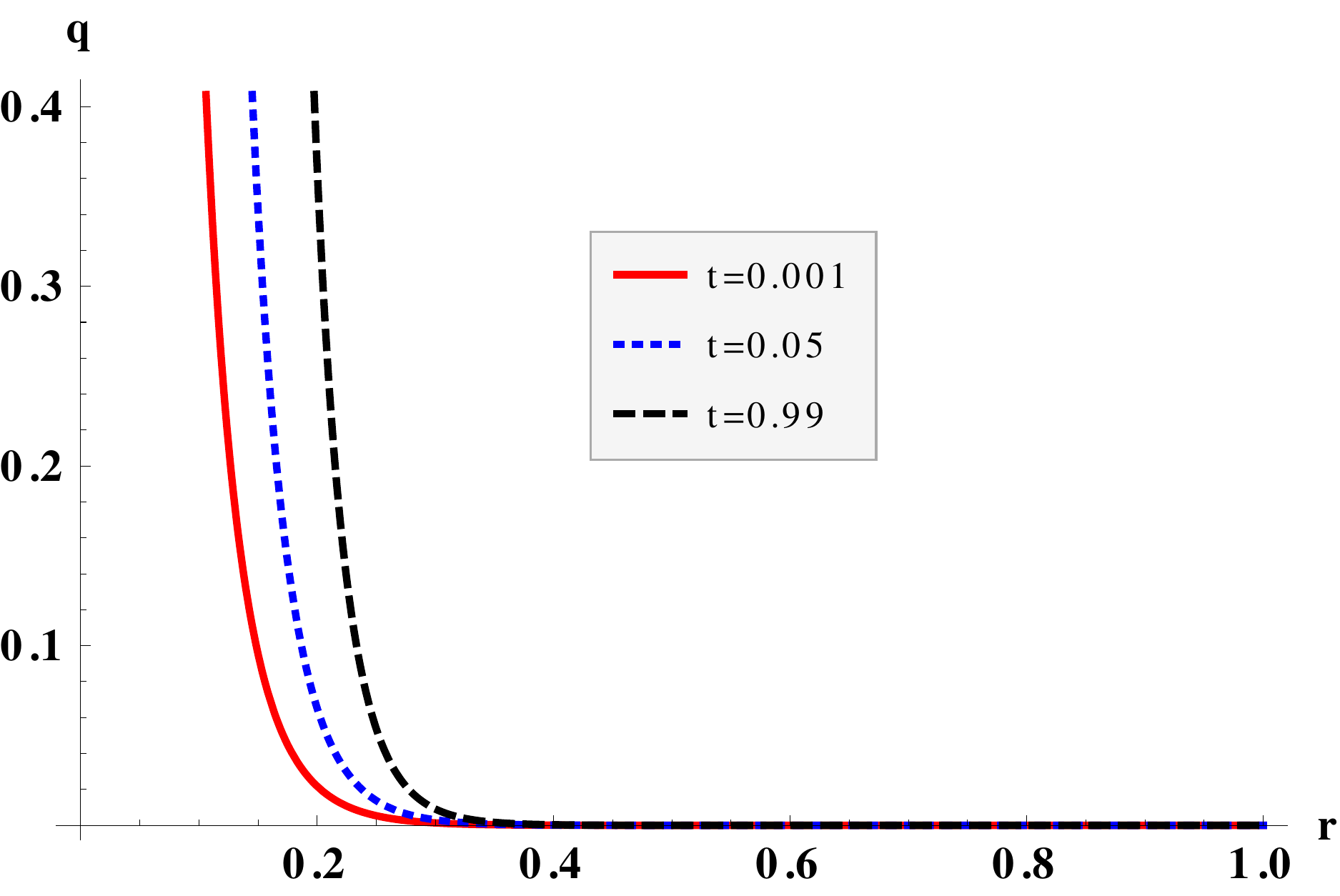}
\caption{$q$ vs $r$}
\label{heatns}
\end{minipage}
\end{figure}
%%%%%%%%%%%%%%%%%%%%%%%%%%%%%%%%%%%%%%%%%%%%%%%%%%%%%%%%%%%%%%
%%%%%%%%%%%%%%%%%%%%%%%%%%%%%%%%%%%%%%%%%%%%%%%%%%%%%%%%%%%%%%%%%
\begin{figure}[t!]
\begin{minipage}[b]{0.3\linewidth}
\centering
\includegraphics[width=2in,height=1.6in]{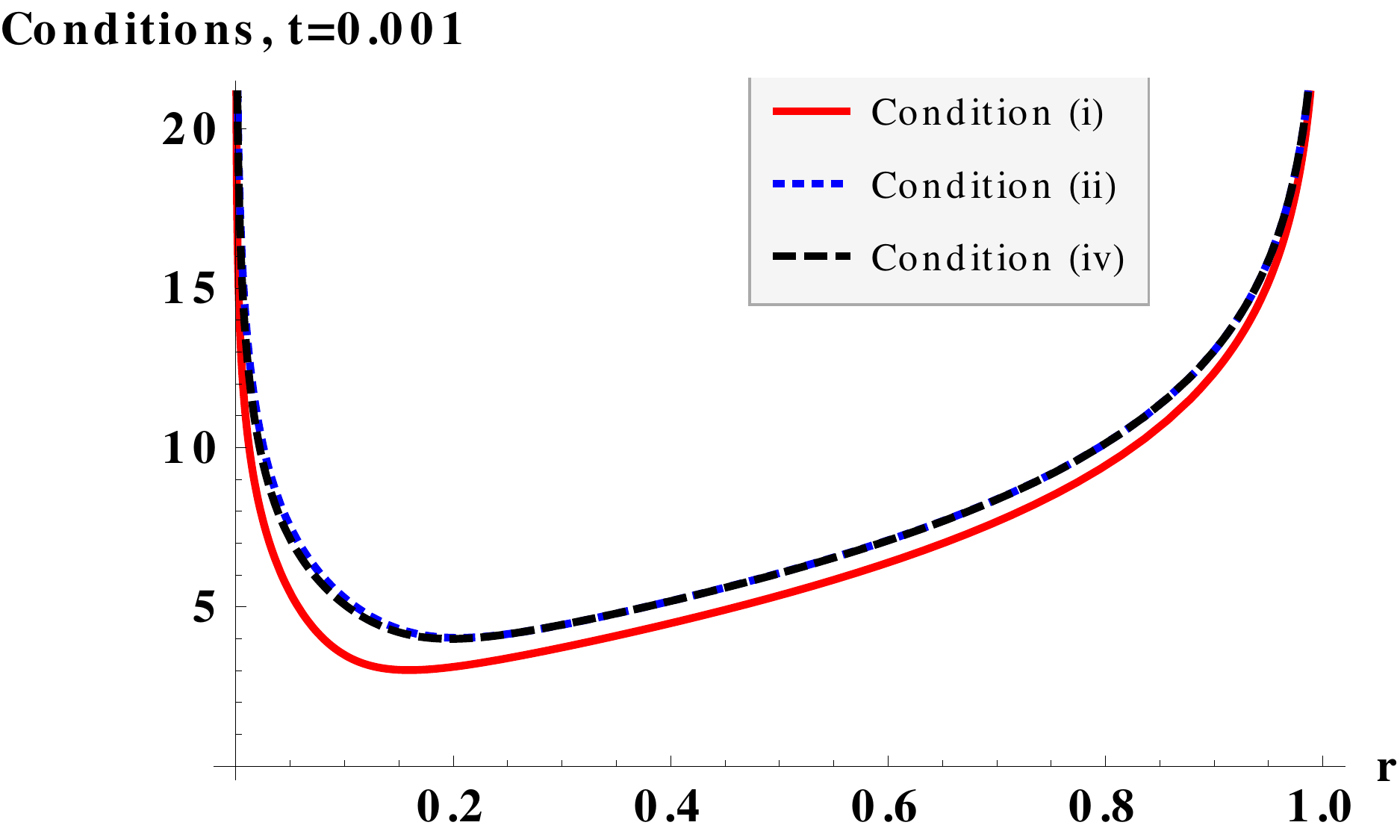}
\caption{Conditions $(i)$, $(ii)$, $(iv)$ at $t=0.001$}
\label{con1ns}
\end{minipage}
\hspace{0.2cm}
\begin{minipage}[b]{0.3\linewidth}
\centering
\includegraphics[width=2in,height=1.6in]{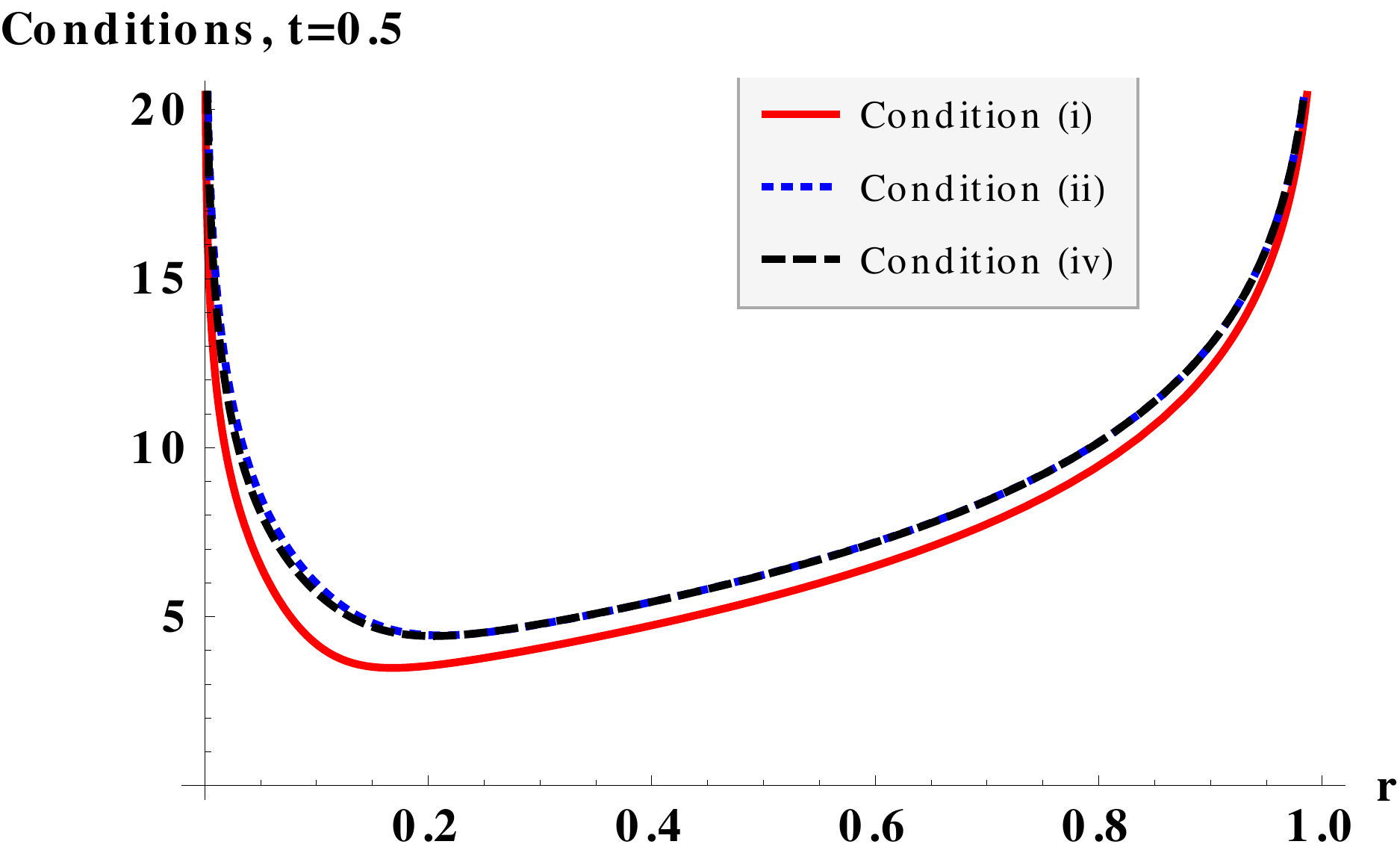}
\caption{Conditions $(i)$, $(ii)$, $(iv)$ at $t=0.5$}
\label{con2ns}
\end{minipage}
\hspace{0.2cm}
\begin{minipage}[b]{0.3\linewidth}
\centering
\includegraphics[width=2in,height=1.6in]{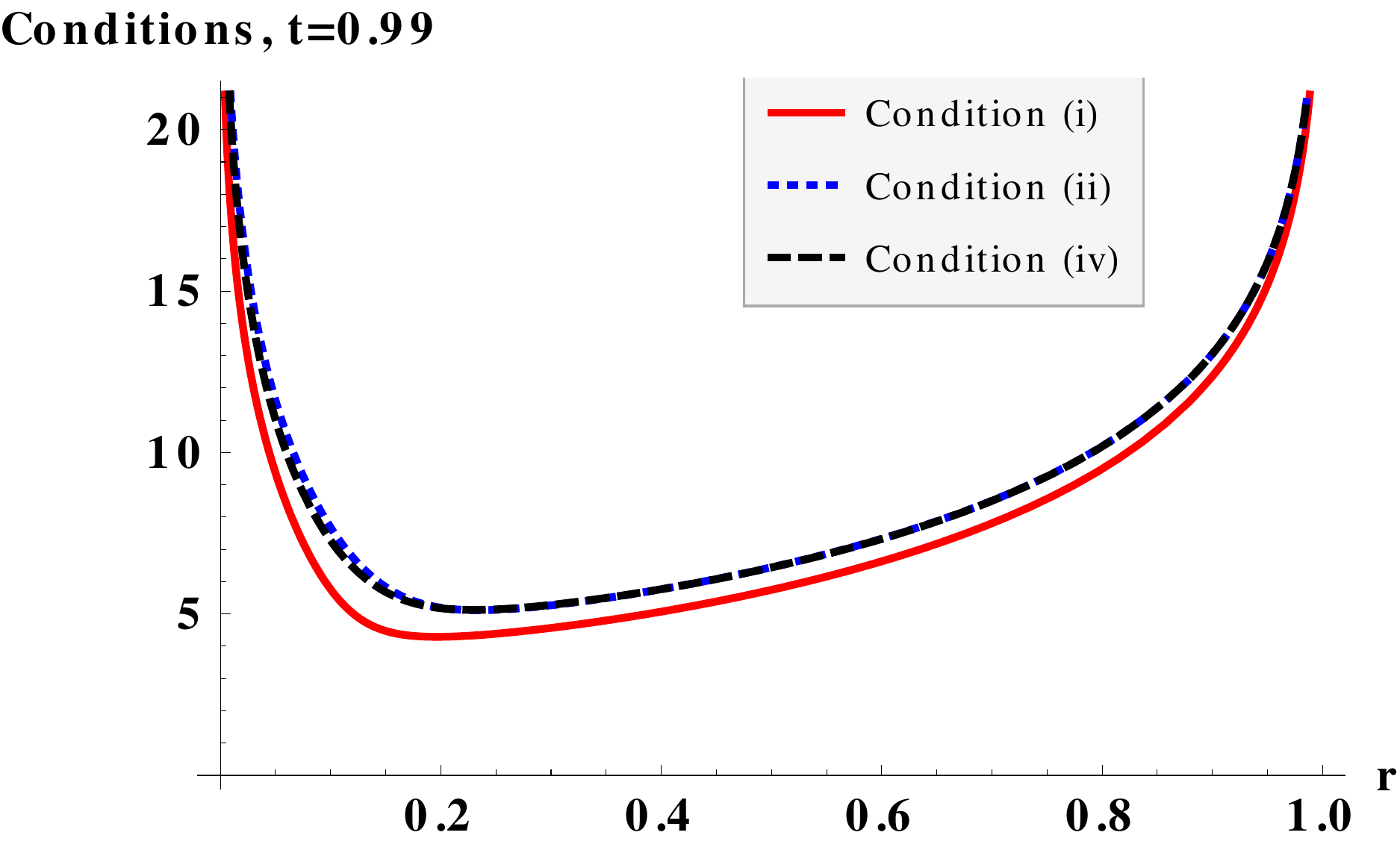}
\caption{Conditions $(i)$, $(ii)$, $(iv)$ at $t=0.99$}
\label{con3ns}
\end{minipage}
\end{figure}
%%%%%%%%%%%%%%%%%%%%%%%%%%%%%%%%%%%%%%%%%%%%%%%%%%%%%%%%%%%%%%
%%%%%%%%%%%%%%%%%%%%%%%%%%%%%%%%%%%%%%%%%%%%%%%%%%%%%%%%%%%%%%%%%
\begin{figure}[t!]
\begin{minipage}[b]{0.3\linewidth}
\centering
\includegraphics[width=2in,height=1.6in]{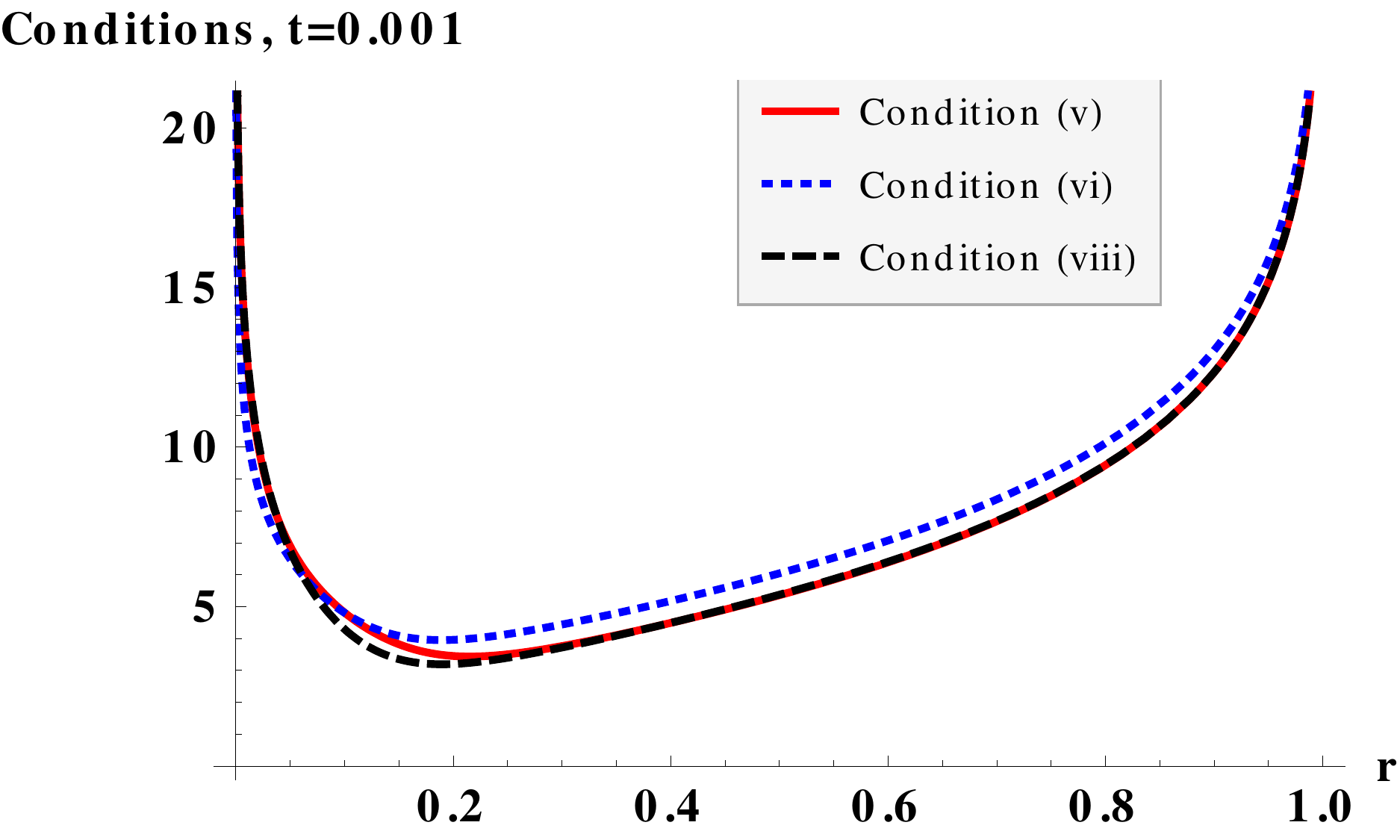}
\caption{Conditions $(v)$, $(vi)$, $(viii)$ at $t=0.001$}
\label{cona1ns}
\end{minipage}
\hspace{0.2cm}
\begin{minipage}[b]{0.3\linewidth}
\centering
\includegraphics[width=2in,height=1.6in]{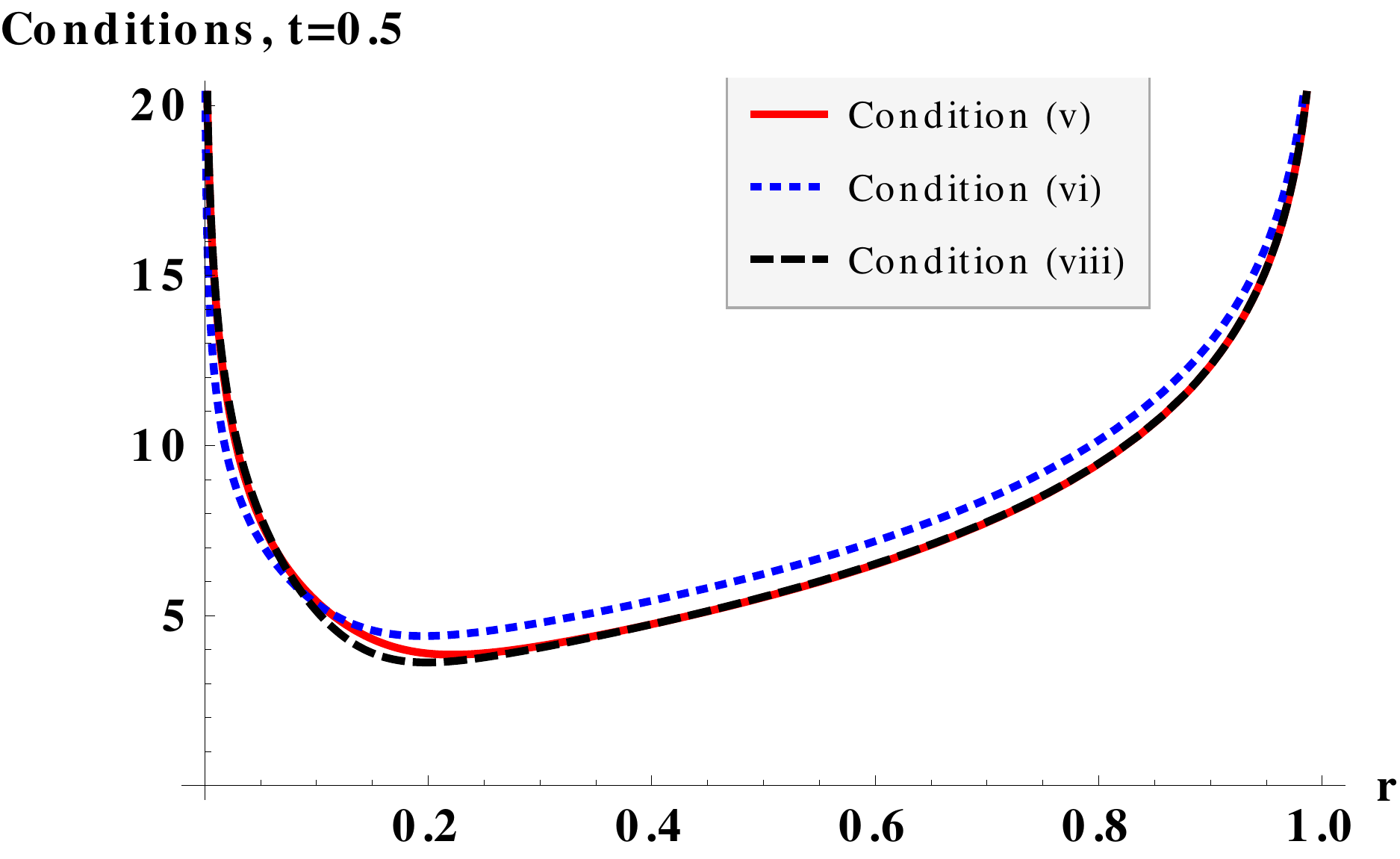}
\caption{Conditions $(v)$, $(vi)$, $(viii)$ at $t=0.5$}
\label{cona2ns}
\end{minipage}
\hspace{0.2cm}
\begin{minipage}[b]{0.3\linewidth}
\centering
\includegraphics[width=2in,height=1.6in]{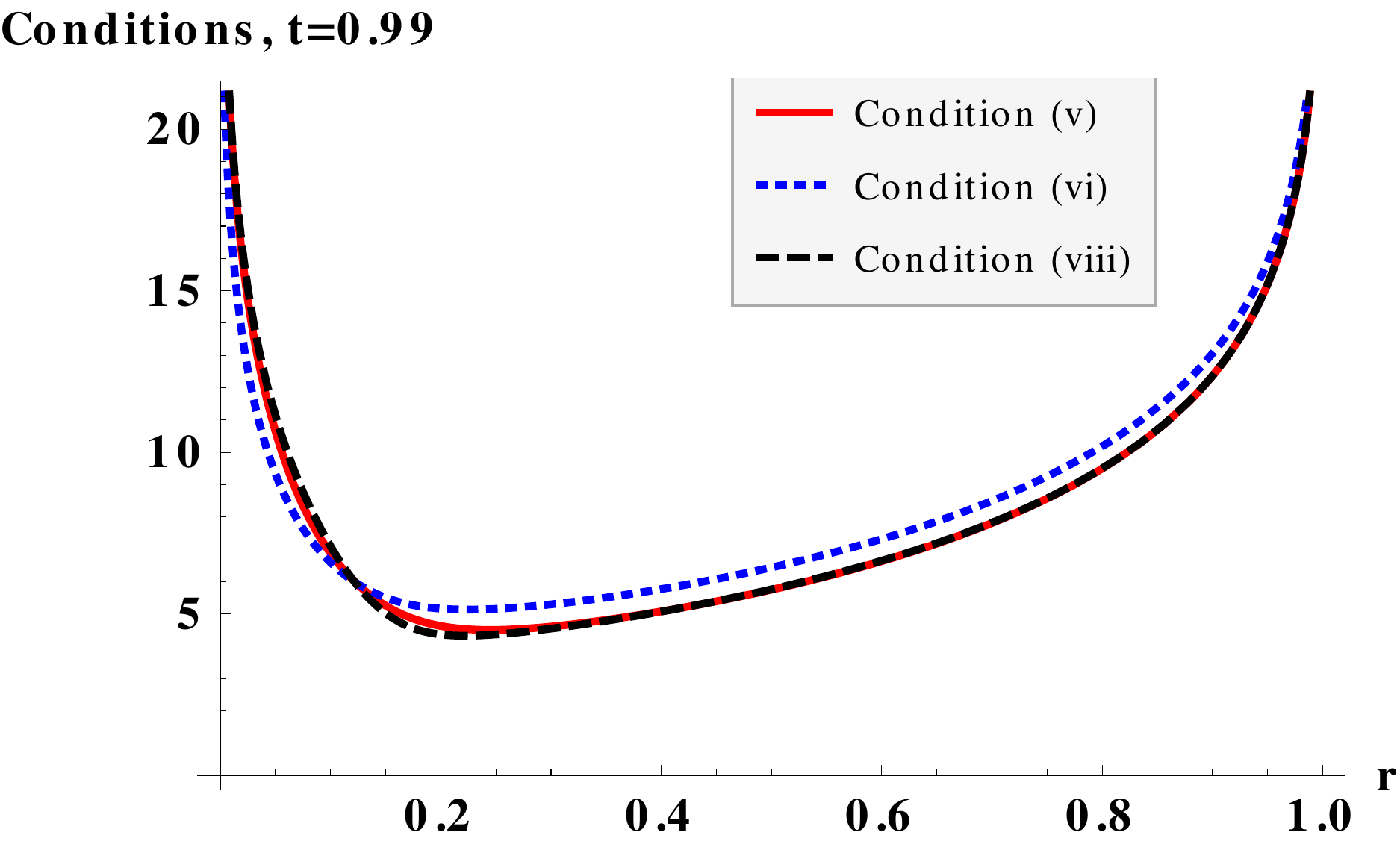}
\caption{Conditions $(v)$, $(vi)$, $(viii)$ at $t=0.99$}
\label{cona3ns}
\end{minipage}
\end{figure}
%%%%%%%%%%%%%%%%%%%%%%%%%%%%%%%%%%%%%%%%%%%%%%%%%%%%%%%%%%%%%%
In figs.(\ref{rhons}), (\ref{anisons}) and (\ref{heatns}), we show the density $\rho$, the anisotropy $\Pi = p_{\theta} - p_r$
and the heat flux $q$ as a function of $r$, for $t=0.001$ (thick red), $0.5$ (dotted blue) and $0.99$ (dashed black), respectively. 
These are computed from eq.(\ref{emt}) and we have set $\eta = 10$ and $\alpha = 10^{-3}$. 
 
It remains to check whether the conditions listed in eqs.(\ref{con12gen}) and (\ref{energyconsgen}) are satisfied during
the collapse process. Without loss of generality, we will make a further choice $\theta = \pi/2$ here, so that 
$\sigma_{22} = \sigma_{33}$ and hence we have to look at the conditions $(i)~,(ii)$ of eq.(\ref{con12gen}) and 
$(iv)~,(v)~,(vi)~$ and $(viii)$ of eq.(\ref{energyconsgen}). Snapshots of the logarithms of the relevant quantities on the left hand
side of the corresponding equations at $t=0.001$, $t=0.05$ and
$t=0.99$ are shown in figs.(\ref{con1ns}) - (\ref{cona3ns}). In figs.(\ref{con1ns}), (\ref{con2ns}) and (\ref{con3ns}), the
thick red, dotted blue and dashed black lines represent the logarithms of conditions $(i)$, $(ii)$ and $(iv)$ and in figs.(\ref{cona1ns}), 
(\ref{cona2ns}) and (\ref{cona3ns}),these represent the logarithms of conditions $(v)$, $(vi)$ and $(viii)$. We find that all the 
required conditions are indeed satisfied. It is also checked that eq.(\ref{finalmod1}) 
is identically satisfied in this case as well. 
As a remark, we note that the shear vanishes at the center (vide eq.(\ref{shear})). Morover, the anistropy diverges at the
origin (due to the singular nature of the solution at $t=0$) at all times during the collapse. 
However, eq.(\ref{finalmod2}) is identically satisfied in this case, as can be checked. 

The solution discussed above collapses into a locally naked singularity, as we have said. We mention in passing that it is also
possible to generate black hole solutions from the generic class of non-separable metrics that we consider here. 
For example, one simply needs to tune the parameter $b$ in eq.(\ref{qrtnonsep}) to $b=2$ (instead of $b=1/2$ used in
the previous example) to see that close to the center, the apparent horizon forms earlier than the singularity (at $t=0.25$).
Again, one can check that all the energy conditions can be satisfied by suitably tuning the parameters $\alpha$ and $\eta$. 
However, we will not go into the details here, as these are entirely similar to the situation that we have considered. 

\section{Nature of the collapsing fluid for separable solutions}

The solutions presented in the previous sections indicate collapse in $f(R)$ gravity to black holes or to singularities that
are locally naked,
while obeying all the energy conditions. A natural question in this context is the physical nature of the fluids, namely if they 
follow an equation of state (EOS). The lack of this analysis is a drawback in many studies of gravitational collapse in $f(R)$
theories that appear in the literature till date. In this context, note that the EOS of collapsing stars is well studied especially
in the non-relativistic limit, starting from the pioneering work of \cite{Bethe}. In the context of $f(R)$ collapse, such a study
is somewhat difficult to envisage, but clearly we can see from eq.(\ref{qtys}) that there is apriori no simple EOS that our
co-moving observer will see, even in the simple case of the separable solutions presented in section 3. We will concentrate
only on this class of solutions in this section, since the solution is section 4 does not correspond to realistic collaps
of a dense star, as already mentioned. 

First, we note that if we set the $f(R)$ parameter $\alpha$ to zero, we have here
\begin{equation}
\frac{p_r}{\rho}\biggr|_{\alpha \to 0} = \frac{p_{\theta}}{\rho} \biggr|_{\alpha \to 0} = 1 - \frac{2}{5-2t}~.
\end{equation}
Hence, at $t=0$, the matter follows a barotropic equation of state with $p_r = p_{\theta} = \gamma \rho$ (remember that
there is no pressure anisotropy with $\alpha \to 0$ as we have commented on at the end of section 3), with $\gamma = 3/5$. 
As the collapse proceeds, the barotropic index reduces in this case, and approaches $1/3$ for $t \to 1$. Hence, at the
end of the collapse, with $\alpha \to 0$, the matter reduces to pure radiation. 

In the general case, the situation is more complicated. Here, we have, from eq.(\ref{qtys}), 
\begin{equation}
\frac{p_r}{\rho}\biggr|_{t \to 0} = \frac{3+28\alpha\left(r-1\right)^4}{5+8\alpha\left(r-1\right)^2}~,~~
\frac{p_{\theta}}{\rho} \biggr|_{t \to 0} = \frac{3-4\alpha\left(r-1\right)^4}{5+8\alpha\left(r-1\right)^2}~,~~
\frac{p_r}{\rho}\biggr|_{t \to 1}=\frac{p_{\theta}}{\rho} \biggr|_{t \to 1} = 1~.
\end{equation}
It is therefore seen that for small values of $\alpha$ (we have used $\alpha = 10^{-3}$ in section 3), 
at the beginning of the collapse, the system is close to a barotropric fluid with $\gamma = 0.6$, but
the effect of $\alpha$ is to increase the barotropic index to unity at the time of formation of the singularity, and at
this time the speed of sound equals the speed of light. However, the latter fact is true strictly at the singularity formation
time, before which the barotropic index is always less that unity. 
We mention in passing that a related question is whether one can envisage a situation where the fluid in question 
consists of two simple fluids, each of which possibly follow an equation of state. This is usually achieved for cases without
shear or heat flow, by rotating the coordinate basis of the co-moving observer. This has been a popular
topic in the literature, starting from the work of \cite{Letelier} (see also \cite{Bayin} for applications in the cosmological
context). It can easily be checked for our model that this is not possible in the presence of heat flux. The intuitive reason
for this is that a dissipative effect cannot be un-done by a rotation of the coordinate basis (unless there is a specific
form of an equation of state which also involves the heat flux, see e.g \cite{KrischGlass2}). 

It is also of interest to consider the heat transport equation in our non-equilibrium collapsing scenario, 
following the pioneering work of \cite{IsraelStewart}. The simplicity
of the solutions derived in eq.(\ref{qtys}) in the separable metric case, allows for explicit computations of the quantities 
appearing in the evolution equation of the heat flux, which reads \cite{IsraelStewart} (see also \cite{Maartenslec}, \cite{Her2})
\begin{equation}
\tau n_{\mu}h^{\mu\nu}q_{\nu;\sigma}u^{\sigma} + q =  -\kappa n_{\mu}h^{\mu\nu}\left(T_{,\nu} + T a_{\nu}\right)
-\frac{1}{2}\kappa T^2 q\left(\frac{\tau u^{\mu}}{\kappa T^2}\right)_{; \mu}~.
\label{tempmain}
\end{equation}
Here, $T$ is the local equilibrium temperature, $\kappa$ is the thermal conductivity, and $\tau$ the relaxation timescale, and
all these quantities must be positive, from physical conditions. Further, $a_{\mu}$ is the acceleration vector defined 
after eq.(\ref{Ray}). 
In order to solve eq.(\ref{tempmain}), a number of assumptions is necessary, since $\kappa$ and $\tau$ are temperature
dependent quantities. There is a vast amount of literature on the topic, and we do not go into the known details here, 
but will simply use the results of 
\cite{GMM}, \cite{GMM2} (see also \cite{GG}) and assume that
\begin{equation}
\kappa = \gamma T^3\tau_c~,~~\tau = \left(\frac{\beta\gamma}{\alpha_1}\right)\tau_c~,~~
\tau_c = \left(\frac{\alpha_1}{\gamma}\right)T^{-\sigma_1}~,
\label{heatqtys}
\end{equation}
where $\beta$, $\gamma$, $\alpha_1$ and $\sigma_1$ are non-negative constants, with the case $\beta = 0$ being the 
non-causal case (see, e.g \cite{Maartenslec} for an excellent exposition). For simplicity, we will restrict ourselves
to cases with $\sigma_1 \leq 4$.

Although eq.(\ref{tempmain}) is in general difficult to solve, the simplicity of the form of the energy-momentum tensor for the 
separable solution considered in section 3 allows us to obtain analytic solutions at least in some approximations. 
First of all, let us consider the non-causal case, and set $\beta = 0$. Then, we obtain the formal solution of eq.(\ref{tempmain}) as
\begin{equation}
T^{4-\sigma_1} = -\frac{\left(1-r\right)^2\left(\sigma_1 - 4\right)}{\alpha_1\left(1-t\right)\left(\sigma_1 - 3\right)}
- \alpha\frac{6\left(1-r\right)^6\left(\sigma_1 - 4\right)}{\alpha_1\left(1-t\right)^2\left(\sigma_1 - 1\right)}
+ \left(1-r\right)^{8-2\sigma_1}F\left(t\right)~~~~~~\left(\beta = 0\right)~,
\label{nc1}
\end{equation}
where $F(t)$ is an apriori undetermined function of the co-moving time. The special cases $\sigma_1 = 1, 3, 4$ need to 
be solved separately. The results are 
\begin{eqnarray}
T^3 &=& -\frac{3\left(1-r\right)^2}{2\alpha_1\left(1-t\right)} + 
\alpha\frac{36\left(1-r\right)^6\log\left(1-r\right)}{\alpha_1\left(1-t\right)^2} + \left(1-r\right)^6 
F(t)~~~~~~\left(\beta=0, ~\sigma_1 =1\right)~,
\nonumber\\
T &=& \left(1-r\right)^2\frac{2\log\left(1-r\right)}{\alpha_1\left(1-t\right)} + \alpha\frac{3\left(1-r\right)^6}{\alpha_1\left(1-t\right)^2}
+ \left(1-r\right)^2F(t)~~~~~~\left(\beta=0, ~\sigma_1 = 3\right)~,\nonumber\\
T &=& {\rm Exp}\left[\frac{\left(1-r\right)^2\left(1-t+2\alpha\left(1-r\right)^4\right)}{\alpha_1\left(1-t\right)^2}\right]
\left(1-r\right)^2 F(t)~~~~~~\left(\beta=0, ~\sigma_1 = 4\right)~,
\label{nc2}
\end{eqnarray}
where we have generically denoted an arbitrary function of the co-moving time by $F(t)$. 
Eqs.(\ref{nc1}) and (\ref{nc2}) are the full set of solutions for the non-causal case, and the role of the $f(R)$ parameter
$\alpha$ can be easily identified, and the 
increase in the core temperature as a function of time is clearly seen. 
In particular, we see from eq.(\ref{nc1}) that the role of $\alpha$ is to decrease the temperature
(compared to the $\alpha = 0$ case) for $\sigma_1 < 1$ and $\sigma_1 > 4$, while it increases the temperature for
$1 < \sigma_1 < 4$. Also, from the first two equations of eq.(\ref{nc2}), it is clear that close to the center,
the effect of $\alpha$ vanishes for $\sigma_1=1$ and dominates for $\sigma_1 = 3$ with the term independent 
of $\alpha$ vanishing in the latter case. No further conclusions can be reached without the knowledge of the arbitrary function $F(t)$.  

However, we can make the following observation from eq.(\ref{nc1}). Close to the boundary, i.e as $r \to 1$,
one can always make the last term on the right hand side of this equation arbitrarily close to zero at a given co-moving
time of the collapse, for $\sigma_1<3$. The fall-off of this term with $r$ being faster than the first term on the right hand side of
eq.(\ref{nc1}) indicates that in such cases, there will exist a domain of the co-moving radius where the temperature 
will not be real (since $\alpha$ is positive). In order to avoid this, we require
$3<\sigma_1<4$ and the other values of $\sigma_1$ are ruled out in the class 
of models that we consider. Note also that the solutions for $\sigma_1 = 3$ and $\sigma_1=4$ do not suffer from
this pathology, and hence our final set of admissible values of $\sigma_1$ is $3\leq \sigma_1 \leq 4$. 

Note that in cases where the interior solution is matched with an external Vaidya metric, the arbitrary
function $F(t)$ can be determined by relating the temperature at the boundary to the luminosity there, and 
then equating this with the luminosity as seen by an observer at infinity, via the red-shift factor. This is not possible
here, as we have matched with an external Schwarzschild solution, for which the temperature and luminosity at
the boundary are automatically zero, as is evident by taking the $r \to 1$ limit in the solutions above. $F(t)$ can thus
be determined in principle if we specify the behaviour of the core temperature as a function of time, along with 
the condition $3 \leq\sigma_1\leq 4$ discussed above.

Finally, we make some comments about the non-causal case. Here, the analysis becomes cumbersome, and 
analytic solutions to the heat flow equation of eq.(\ref{tempmain}) seem difficult to obtain. As a somewhat crude 
approximation (used in \cite{GMM}, \cite{GMM2}, \cite{GG}), if we ignore the last term on the right hand side of eq.(\ref{tempmain}),
then with eq.(\ref{heatqtys}), we obtain as a solution for $\sigma_1 = 0$, 
\begin{equation}
T^4 = -\frac{4\left(1-r\right)^2\left(1-t\right) + 9\beta\left(1-r\right)^4}{3\alpha_1\left(1-t\right)^2} +
\alpha\frac{24\left(1-r\right)^6\left[t-1+5\beta\left(1-r\right)^2\log\left(1-r\right)\right]}{\alpha_1\left(1-t\right)^3} + \left(1-r\right)^8F(t)~,
\label{causal}
\end{equation}
where again the arbitrary function of time can be constrained if we assume a time profile of the core temperature. 
 
The analysis in this section was related to the separable solutions that we have used in section 3. For the non-separable
solutions of section 4, such analyses become quite tedious, and will not provide much physical insight, as should be evident
from the comments made at the beginning of that section.

\section{Discussions}

Gravitational collapse in metric $f(R)$ theories of gravity are greatly restricted due to the extra junction conditions that
have to be invoked, and involve the continuity of the Ricci scalar and its derivatives across a time-like hypersurface on which
an internal collapsing metric is matched with an external solution. This is the $R$-matching method commonly used
in $f(R)$ scenarios. In fact, there are a total of twelve conditions that will generically
need to be satisfied, and are given in eqs.(\ref{allmatch}), (\ref{con12gen}) and (\ref{energyconsgen}). It is indeed a formidable
task to compute explicit collapsing solutions while satisfying all these conditions and not many exact solutions are
available in the literature. 

In this paper, we have constructed novel examples of $f(R)$ collapse, by using the extra junction conditions to 
explicitly solve for the collapsing metric. This was done in $f(R) = \alpha R^2$ theories of gravity in two cases, one in which 
we assumed a separable form of the internal metric, and the other in which this assumption was relaxed. 
We showed that by suitably choosing some reasonable forms of a few 
arbitrary functions, new examples of collapse in modified gravity can be constructed. In the
latter context, we have described a collapse situation that includes the effects of shear viscosity.
The generic relation between the shear viscosity and the anisotropy in our $f(R)$ models has been derived here. 
We have demonstrated by explicit examples the formation of black holes or naked
singularities, while satisfying all the energy conditions. 

The separable solution constructed by us allows for analytical forms of the components of the energy momentum tensor.
Using these, we are also able to obtain analytical solutions to the evolution equation of the heat flux. Here, the simplicity
of the expressions involved allows us to focus on the effect of the $f(R)$ parameter $\alpha$, with certain reasonable
approximations. As mentioned in the text, this situation corresponds to a realistic collapse of a dense star, with the
pressures remaining positive at all times. It would be interesting to understand the evolution of the entropy in this
non-equilibrium situation. 

Our analysis in this paper relies on a number of explicit choices that we have made, and these have been highlighted
in sections 3 and 4. Indeed, these choices are arbitrary and serve as examples of more general cases than what we
study here, and functions different from what we have chosen should generate more physical examples of collapse
scenarios in modified gravity. Further, our analysis here is limited to models with $f(R) = R^2$. 
It should be interesting to apply this to more generic situations. 

\begin{center}
{\bf Acknowledgements}
\end{center}

TS thanks Pratim Roy for useful discussions. The work of SC is supported by CSIR, India, 
via grant number 09/092(0930)/2015-EMR-I.

\end{document}